\documentclass[aps,prd,notitlepage,showpacs,nofootinbib,superscriptaddress]{revtex4-1}

\usepackage{graphicx}
\usepackage[utf8]{inputenc} 

\usepackage{amsmath}
\usepackage{amssymb}
\usepackage{float}
\usepackage{comment}
\usepackage{slashed}
\usepackage[normalem]{ulem}

   \makeatletter
     \renewcommand\@make@capt@title[2]{%
      \@ifx@empty\float@link{\@firstofone}{\expandafter\href\expandafter{\float@link}}%
       {\textbf{#1}}\@caption@fignum@sep#2\quad}%
     \makeatother

\makeatletter 
\renewcommand{\fnum@figure}{\textbf{Fig.~\thefigure}}
\makeatother
\usepackage{multirow}
\usepackage{rotating}

\usepackage[usenames,dvipsnames]{color}
\usepackage[colorinlistoftodos]{todonotes}
\usepackage[colorlinks=true,citecolor=darkred,urlcolor=darkred, pdfborder={0 0 0}]{hyperref}
\usepackage[normalem]{ulem}

\definecolor{darkred}{rgb}{0.6,0,0}
\usepackage[colorinlistoftodos]{todonotes}

\definecolor{linkcolor}{rgb}{0,0,0.5}

\newcommand {\ignore}[1]{}

\def\gsim{\raise0.3ex\hbox{$\;>$\kern-0.75em\raise-1.1ex\hbox{$\sim\;$}}}
\def\lsim{\raise0.3ex\hbox{$\;<$\kern-0.75em\raise-1.1ex\hbox{$\sim\;$}}}

\providecommand{\be}{ \begin{equation} } 
\providecommand{\ee}{ \end{equation} }
\providecommand{\bea}{\begin{eqnarray}}
\providecommand{\eea}{\end{eqnarray}}
\providecommand{\nn}{\nonumber}

\providecommand{\to}{\rightarrow}

\usepackage{soul}

\definecolor{mightnightblue}{RGB}{25,25,112}

\definecolor{brown}{rgb}{0.59, 0.29, 0.0}

\newcommand {\black} {\color{black}}

\def\vev#1{\left\langle #1\right\rangle}

\def\21{$\mathrm{SU(2)_L \otimes U(1)_Y}$}

\def\3311{$\mathrm{SU(3) \otimes SU(3)_L \otimes U(1)_X \otimes U(1)_{N}}$ }

\newcommand{\AddrTHU}{Center of High Energy Physics, Tsinghua University, Beijing 100084, China.}
\newcommand{\AddrAHEP}{%
  AHEP Group, Institut de F\'{i}sica Corpuscular --
  C.S.I.C./Universitat de Val\`{e}ncia, Parc Cient\'ific de Paterna.\\
 C/ Catedr\'atico Jos\'e Beltr\'an, 2 E-46980 Paterna (Valencia) - Spain}
\newcommand{\AddrUCN}{Departamento de Física, Universidad Católica del Norte, Avenida Angamos 0610, Casilla 1280, Antofagasta, Chile.}
\newcommand{\AddrUFABC}{Centro de Ci\^encias Naturais e Humanas, Universidade Federal do ABC, 09210-580, Santo Andr\'e-SP, Brasil}

\begin{document}

 \title{\boldmath \color{BrickRed} Phenomenology of fermion dark matter as neutrino mass mediator with gauged B-L}

\author{Carlos Alvarado}\email{arcarlos00@gmail.com}
\affiliation{\AddrTHU}

\author{Cesar Bonilla} \email{cesar.bonilla@ucn.cl}
\affiliation{\AddrUCN}

\author{ Julio Leite }\email{julio.leite@ufabc.edu.br}
\affiliation{\AddrUFABC}

\author{Jos\'{e} W. F. Valle}\email{valle@ific.uv.es}
\affiliation{\AddrAHEP}

\begin{abstract}
\vspace{0.5cm}

We analyze a model with unbroken $U(1)_{B-L}$ gauge symmetry where neutrino masses are generated at one loop, after spontaneous breaking of a global $U(1)_{G}$ symmetry.
These symmetries ensure dark matter (DM) stability and the Diracness of neutrinos. Within this context, we examine fermionic dark matter.
Consistency between the required neutrino mass and the observed relic abundance indicates dark matter masses and couplings within the reach of direct detection experiments.
\end{abstract}

\maketitle
\noindent 

\section{Introduction}

Two of the major drawbacks of the Standard Model (SM) is the absence of neutrino masses and of a viable dark matter candidate, for both of which we have strong evidence.
Neutrino masses are clearly required in order to account for the neutrino oscillation data~\cite{deSalas:2020pgw}, while the existence of dark matter (DM) is strongly supported by observational
evidence at multiple scales through gravitational effects.
These include the role of DM in structure formation as well as its influence on the Cosmic Microwave Background (CMB).
Dark matter constitutes about 80\% of the matter content of the Universe.
CMB studies by the PLANCK collaboration yield the following value for the dark matter relic abundance~\cite{Aghanim:2018eyx},
\begin{equation}
\Omega_{\text{DM}}h^{2}=0.120\pm0.001~~~~~\text{at}~~~~~90\%\text{ C.L}~.
\label{eq:omega}
\end{equation}

No one knows the origin of neutrino mass nor the nature of dark matter.
It seems likely that DM is a weakly interacting massive particle (WIMP), stable on cosmological time scales.

It has been suggested that neutrino mass generation and dark matter are closely interconnected.
In this letter we explore the specially interesting \textit{scotogenic} possibility that dark matter mediates neutrino mass generation~\cite{Ma:2006km,Hirsch:2013ola,Merle:2016scw,Avila:2019hhv,Boehm:2006mi,Farzan:2009ji}.

Gauge extensions of the Standard Model provide an interesting setting to examine the interconnection between neutrino masses and the properties of dark matter
candidates~\cite{Farzan:2012sa,Dong:2018aak,Kang:2019sab,Leite:2019grf,CarcamoHernandez:2020ehn,Calle:2019mxn}.
We do so within minimal U(1) gauge extensions of the Standard Model. 
We consider the case where an exact local $U(1)_{B-L}$ symmetry is responsible for the stability of dark matter,
while neutrino masses arise radiatively thanks to the spontaneous breaking of a $U(1)_G$ global symmetry.
This is in contrast with Ref.~\cite{Reig:2018mdk} which considered the case of a global $U(1)_{B-L}$ symmetry. 
Moreover, here we have \emph{elementary}, rather than \emph{bound-state} dark matter considered in~\cite{Reig:2018mdk}.
Our $U(1)_{B-L}$ gauge symmetry is conserved but thanks to the Stueckelberg mechanism~\cite{Leite:2020wjl} the associated gauge boson becomes massive, 
while the spontaneous breaking of $U(1)_G$ is responsible for generating neutrino masses. 
The latter implies the existence of a physical Nambu-Goldstone boson, the \textit{Diracon}~\cite{Bonilla:2016zef,Bonilla:2016diq}.
Strict ${B-L}$ conservation implies that neutrinos should be Dirac fermions, while
the requirement of generating viable neutrino masses that can account for the neutrino oscillation data restricts fermionic dark matter masses and couplings to regions that can be probed in
upcoming nuclear recoil scattering experiments.

In the next section we present the charge assignments and mass spectrum of our model. Section \ref{sec:constraints} provides the relevant constraints for our analysis.
The results of our numerical study are described in Section~\ref{sec:FermionDM}, and a summary and outlook are given in Section \ref{sec:conclusions}.

\section{The model}
\label{sec:model}

 We propose a SM extension based on the $\mathrm{SM}\otimes U(1)_{B-L}\otimes U(1)_G$ symmetry.
   The $U(1)_{B-L}$ symmetry is local and fully conserved, while the $U(1)_G$ is global and spontaneously broken.
  The fermion sector of our model is extended with respect to that of the SM by \textit{right-handed neutrinos}, $\nu_{R}$, and vectorlike pairs $S_{L}$, $S_{R}$.
 As for the scalar sector, in addition to the SM Higgs doublet, $H$, we introduce another $SU(2)_L$ doublet $\eta$ as well as two singlets $\xi$ and $\sigma$.
  As shown in detail in what follows, the new fields are crucial for neutrino mass generation and dark matter phenomenology.
The lepton and scalar content and corresponding symmetry transformations are given in Table~\ref{Tab1}. 

 The scalars charged under $U(1)_{B-L}$, i.e. $\eta$ and $\sigma$, do not acquire a vacuum expectation value (vev), ensuring $B-L$ conservation.
In contrast, the standard Higgs mechanism takes place in the $SU(2)_L \otimes U(1)_Y$ sector when the scalar doublet $H$ acquires a vev.
Meanwhile, when $\xi$ acquires a vev, the global $U(1)_G$ symmetry is spontaneously broken giving rise to a Goldstone boson, dubbed Diracon~\cite{Bonilla:2016zef,Bonilla:2016diq}.
The Diracon is analogous to the \textit{Majoron} that appears when Majorana masses arise following the spontaneous breaking of the global lepton number symmetry.
Being a gauge singlet, its main observational effects would come from the Higgs sector and cosmology.
\begin{table}[h!]
\begin{center}
\begin{tabular}{| c || c | c | c | c |}
  \hline 
&   Fields            &    $SU(2)_L \otimes U(1)_Y$            &     $U(1)_{B-L}$   &     $U(1)_G$               \\
\hline \hline
\multirow{4}{*}{ \begin{turn}{90}\,Leptons \end{turn} } &
 $L_{L}$        	  &    ($\mathbf{2}, {-1/2}$)       &   ${-1}$ & ${0}$                         \\	
 &$e_{R}$     &   ($\mathbf{1}, {-1}$)      & $-1$ &  ${0}$ \\
&   $\nu_{R}$       &   ($\mathbf{1}, {0}$)      & $-1$ &  ${-1}$ \\
&  $S_{L,R}$     	  &  ($\mathbf{1}, {0}$) 	     & $2n$ &  ${0}$      \\
\hline \hline
\multirow{4}{*}{ \begin{turn}{90} Scalars \end{turn} } &
  $H$  		 &  ($\mathbf{2}, {1/2}$)      &  ${0}$     &  ${0}$    \\
  & $\xi$             &  ($\mathbf{1}, {0}$)        &  ${0}$  &  ${1}$   \\	
& $\eta$          	 &  ($\mathbf{2}, {1/2}$)      &  ${2n+1}$ &  ${0}$ \\
& $\sigma$          	 &  ($\mathbf{1}, {0}$)        &  ${2n+1}$    &  ${1}$    \\		
    \hline
  \end{tabular}
\end{center}
\caption{Field content and charge assignments: $B-L$ (gauged, unbroken) and $G$ (global, spontaneously broken). }
 \label{Tab1} 
\end{table}

The exact conservation of the $B-L$ symmetry implies that the matter-parity subgroup, defined as 
\be\label{MP}
M_P = (-1)^{3(B-L)+2s},
\ee
also remains unbroken. 
Under $M_P$, all the SM fields as well as $\nu_R$ and $\xi$ transform trivially, while $S_L,S_R,\eta$ and $\sigma$ are all $M_P$-odd. 
Therefore, the lightest among the $M_P$-odd fields is stable and, if electrically neutral, can play the role of WIMP dark matter. For definiteness, $n=1$ is adopted in the rest of the paper. 

\subsection*{Scalar sector and symmetry breaking}
\label{sec:Sca}

The scalar potential can be written as
\begin{eqnarray}
 V&=&  - \mu_H^2 H^\dagger H + \mu_\eta^2 \eta^\dagger \eta -\mu_\xi^2 \xi^*\xi +\mu_\sigma^2 \sigma^*\sigma
 + \lambda_H (H^\dagger H)^2 + \lambda_\eta (\eta^\dagger \eta)^2+\lambda_\xi (\xi^{*}\xi)^2 
+\lambda_\sigma (\sigma^{*}\sigma)^2\notag\\
&+& \lambda_{\sigma H} (\sigma^{*}\sigma)(H^\dagger H)
+ \lambda_{\sigma \xi} (\sigma^{*}\sigma)(\xi^{*} \xi)
+ \lambda_{\sigma \eta} (\sigma^{*}\sigma)(\eta^\dagger \eta)
+ \lambda_{H \xi} (H^\dagger H)(\xi^{*} \xi)\notag\\
&+& \lambda_{H \eta} (H^\dagger H)(\eta^{\dagger} \eta)+\lambda'_{H \eta} (H^\dagger \eta)(\eta^\dagger H)
+ \lambda_{\xi\eta} (\xi^{*} \xi)(\eta^\dagger \eta)
+\lambda_{D} (\eta^\dagger H \sigma \xi^{*} + \text{h.c}.).\label{eq:V}
\end{eqnarray}

Out of the four scalar fields, two are Higgs doublets, $H= (H^+, H^0)^T$ and $\eta= (\eta^+, \eta^0)^T$, while two are singlets, $\xi$ and $\sigma$.
We assume that only $H$ and $\xi$ acquire vevs $ v$ and $v_\xi$, as follows, 
\begin{equation}
 H^0 = \frac{1}{\sqrt{2}}(v + S_H + iA_{H}), \quad \quad  \xi = \frac{1}{\sqrt{2}}(v_\xi + S_\xi + iA_{\xi}).
\end{equation}
The tadpole equations that follow from the potential are
\begin{eqnarray}
 v \left(\mu_H^2+\lambda_H v^2+ \frac{\lambda_{H\xi} v_\xi^2}{2}\right)&=& 0\,,\nonumber\\
 v_\xi \left(\mu_\xi^2 + \lambda_\xi  v_\xi^2+ \frac{\lambda_{H\xi} v^2}{2}  \right)&=& 0\,,
\end{eqnarray}
which we solve for $\mu_{H}^{2}$ and $\mu_{\xi}^{2}$. 
In order to determine the scalar spectrum, we consider first the CP- and $M_P$-even fields, in the basis $(S_H, S_\xi)$, and write the associated squared mass matrix as 
\begin{eqnarray}
\mathcal{M}_{S}^{2}=
\left(
\begin{array}{cc}
 2 \lambda_H v^2 & \lambda_{H\xi} v v_\xi \\
 \lambda_{H\xi} v v_\xi & 2 \lambda_\xi  v_\xi^2  \\
\end{array}
\right)~.
\end{eqnarray}
Changing to the physical or mass basis 
\begin{eqnarray}
 H_{1} &=& \cos \theta_{h} S_H +\sin \theta_{h} S_\xi\,\nonumber\\
 H_{2} &=& -\sin \theta_{h} S_H + \cos \theta_{h} S_\xi~,
\end{eqnarray}
where 
\begin{equation}
  \tan (2\theta_h) = \frac{v v_\xi \lambda_{H\xi} }{v^2 \lambda_H -v_\xi^2 \lambda_\xi },
\end{equation}
we find a diagonal matrix with eigenvalues given by
\begin{equation}
m_{H_{1,2}}^2= \lambda_H v^2+\lambda_\xi  v_\xi^2 \mp \sqrt{\lambda_H^2 v^4+v^2 v_\xi^2 \left(\lambda_{H\xi}^2 - 2 \lambda_H \lambda_\xi \right)+\lambda_\xi^2 v_\xi^4}\,.
\end{equation}
One sees that, in the limit $v_\xi/v\gg 1 $, the mixing angle becomes very suppressed, so that
$H_{1} \simeq S_H$ where $m_{H_{1}}^2 \simeq v^2\left(2 \lambda_H - \lambda_{H\xi}^2/(2\lambda_\xi)\right)$ is identified with the $125$ GeV Higgs boson,
while $H_2$ is a heavier neutral scalar, with mass $m_{H_{2}}^2 \simeq 2 \lambda_\xi v_\xi^2$ and mainly composed by the singlet $S_{\xi}$. \\[-.3cm]

The CP-odd scalars, $A_{H}$ and $A_{\xi}$, remain unmixed and massless.
The first one is absorbed by the $Z$ boson, whereas $A_{\xi}$ is the physical \textit{true} Goldstone boson associated with the spontaneous breaking of the global $U(1)_G$ symmetry. 
This is closely related to the radiative Dirac neutrino mass generation, so it will be referred to as the Diracon, $\mathcal{D}\equiv A_{\xi}$ \cite{Bonilla:2016zef,Bonilla:2016diq}.

The electrically charged fields $H^\pm$ and $\eta^\pm$ do not mix since they transform differently under the $B-L$ symmetry.  
The former is absorbed by the gauge sector, while the latter, the dark charged scalar, gets the following mass 
\be \label{ChScM}
m_{\eta^\pm}^2 = \frac{1}{2}\left(\lambda_{H\eta} v^2+\lambda_{\xi\eta} v_\xi^2\right) + \mu_{\eta}^2~.
\ee

Finally, the complex neutral fields $(\sigma,\,\eta^0 )$ in the ``dark sector'' mix after spontaneous symmetry breaking according to the squared mass matrix 
\be \label{CNScMM}
\mathcal{M}^{2}_{\varphi}= \frac{1}{2}\left(
\begin{array}{cc}
 2\mu_\sigma^2 + \lambda_{H\sigma} v^2 + \lambda_{\sigma\xi} v_\xi^2  & \lambda_{D} v v_\xi \\ \lambda_{D} v v_\xi  & 2\mu_\eta^2 + (\lambda_{H\eta}+\lambda^\prime_{H\eta})  v^2 + \lambda_{\xi\eta} v_\xi^2 \\
\end{array}
\right)\,.
\ee
After diagonalisation, we find two massive scalars 
\bea \label{CNSc}
\begin{pmatrix} \varphi_{1}^0 \\ \varphi_{2}^0 \end{pmatrix} = \begin{pmatrix} \cos \theta_\varphi & \sin \theta_\varphi\\ -\sin \theta_\varphi & \cos \theta_\varphi \end{pmatrix} \begin{pmatrix} \sigma\\ \eta^0 \end{pmatrix},\,\mbox{with}\, \tan(2\theta_\varphi) = \left[\frac{2 \lambda_{D} v v_\xi}{2(\mu_\sigma^2-\mu_\eta^2) + (\lambda_{H\sigma}-\lambda_{H\eta}-\lambda^\prime_{H\eta})v^2+(\lambda_{\sigma\xi}-\lambda_{\xi\eta})v_\xi^2 }\right],
\eea 
whose masses are
\bea \label{CNScM}
m_{\varphi^0_{1,2}}^2 &=& \frac{1}{4}\bigg\{2 (\mu_\eta^2 + \mu_\sigma^2)+ (\lambda_{H\eta} + \lambda^\prime_{H\eta} + \lambda_{H\sigma})v^2 +(\lambda_{\sigma\xi}+\lambda_{\xi\eta})v_\xi^2 \notag \\
&&\mp \mathcal{F} \sqrt{\left[2(\mu_\sigma^2-\mu_\eta^2 ) + ( \lambda_{H\sigma}-\lambda_{H\eta} - \lambda^\prime_{H\eta})v^2+(\lambda_{\sigma\xi}-\lambda_{\xi\eta})v_\xi^2\right]^2+4\lambda_{D}^2 v^2 v_\xi^2} \bigg\}\,,
\eea
respectively. Here $\mathcal{F}=1$ for $(\mathcal{M}_{\varphi}^{2})_{22}/(\mathcal{M}_{\varphi}^{2})_{11}>1$ and $\mathcal{F}=-1$ otherwise. 

\subsection*{Dirac neutrino mass}
\label{subsec:dnu}

\begin{figure}[h!]
\centering
\includegraphics[scale=0.6]{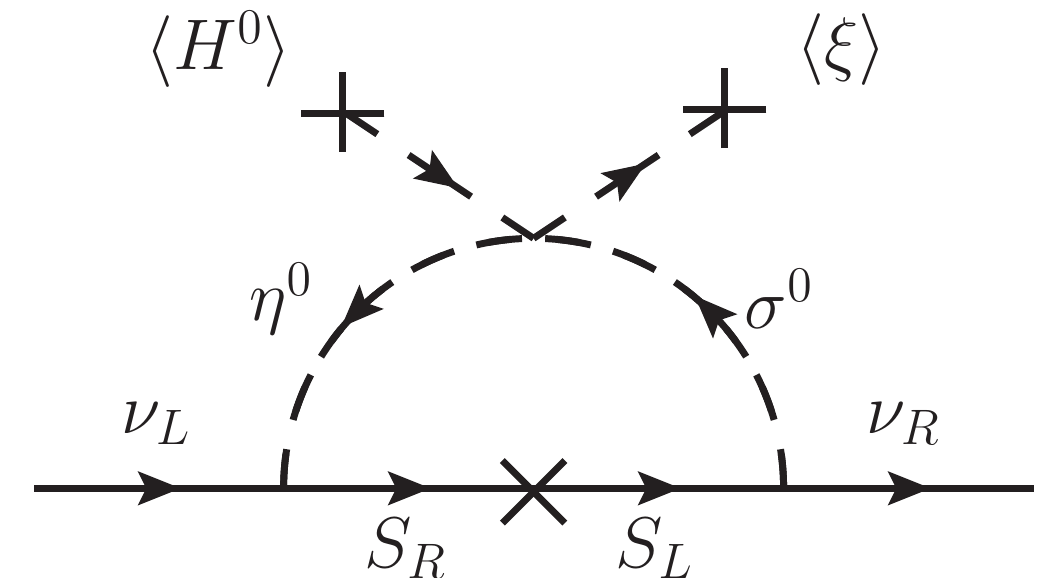}
\caption{Scotogenic Dirac neutrino mass generation, in the interaction basis.}
\label{fig:ScotoLoop}
\end{figure}

The Yukawa Lagrangian for leptons is given by\footnote{The Yukawa Lagrangian for quarks is omitted for it is identical to the SM case.} 
\begin{equation}
-\mathcal{L}_Y = y^e~\overline{L_L} H e_R + y^\nu~\overline{L_L} \widetilde{\eta} S_R + y^{\sigma}~\overline{S_L} \sigma \nu_R + M_S^D\overline{S_L}S_R + \text{h.c}.~,\label{eq:Yuk}
\end{equation}
where the flavor indices have been omitted. The first term is responsible for charged lepton masses when $H$ acquires a vev, as usual.
Notice that the term $\overline{L_L} \widetilde{H} \nu_R$ is forbidden by the $U(1)_G$ symmetry. 
Nonetheless, when $U(1)_G$ is spontaneously broken by $\vev{ \xi}\neq0$, neutrino masses are generated radiatively through the one-loop diagram in Fig.~\ref{fig:ScotoLoop}.

The \textit{dark} fields mediating neutrino masses in Fig. \ref{fig:ScotoLoop} are odd under $M_P$, defined in Eq. (\ref{MP}), and hence the lightest amongst them is stable and can play the role of dark matter.
In the mass basis, the scalar mediators $\eta^0$, $\sigma$ become $\varphi_{1,2}$, according to Eq. (\ref{CNSc}).
From the Yukawa Lagrangian in Eq. (\ref{eq:Yuk}) one sees that the vectorlike $S_{L}$ and $S_{R}$ mediators have bare Dirac masses, determined by diagonalizing $M_{S}^{D\dag}M_{S}^{D}$.
Without loss of generality $M_{S}^{D}$ can be assumed diagonal and the (increasingly ordered) $S_{k}$ physical Dirac masses are  
\begin{equation}
m_{Sk}=(M_{S}^{D})_{kk},~~~~~k=1,2,3~. \label{eq:massS}
\end{equation}

We now turn to the neutrino mass matrix which results from the loop in Fig. \ref{fig:ScotoLoop}.
In the mass basis of the $S_{j}$~\footnote{In order to ensure a rank-2 neutrino mass matrix consistent with neutrino oscillation data we require extra dark mediators.}
and $\varphi_{i}$ one gets, 
\begin{equation}
(m_{\nu})_{ij}=\dfrac{\sin{(2\theta_{\varphi})}}{32\pi^{2}}\sum_{k}y_{ik}^{\nu}y_{kj}^{\sigma}m_{S_{k}}\left[ \dfrac{m_{\varphi_{1}}^{2}}{m_{\varphi_{1}}^{2}-m_{S_{k}}^{2}}\log{\dfrac{m_{\varphi_{1}}^{2}}{m_{S_{k}}^{2}}}-\dfrac{m_{\varphi_{2}}^{2}}{m_{\varphi_{2}}^{2}-m_{S_{k}}^{2}}\log{\dfrac{m_{\varphi_{2}}^{2}}{m_{S_{k}}^{2}}} \right]~. \label{eq:nuLoopMass}
\end{equation}
It is crucial for nonzero neutrino mass that $\lambda_{D}\neq0$ and hence $\theta_{\varphi}\neq0$. 
Notice that $\lambda_{D}\neq0$ is equivalent to nondegeneracy between the scalars running in the loop, $m_{\varphi_{1}}\neq m_{\varphi_{2}}$, rather than nondegeneracy between the real and imaginary parts.
That is, $m_{\varphi iR}=m_{\varphi iI}$ holds regardless of the $\lambda_{D}$ value. 
Indeed, notice that the scalar vertex in the neutrino mass loop follows from the $U(1)_{G}-$invariant operator $\eta^{\dag}H\sigma\xi^{*}$, instead of a soft-breaking term as in
  Ref. \cite{Leite:2020wjl} and other scotogenic constructions. 
 Here instead, the smallness of $m_{\nu}$ is associated with the spontaneous breaking of the global $U(1)_{G}$ symmetry through the vev $v_{\xi}\neq 0$.
  Indeed, $m_{\nu} \to 0$ as $\lambda_D \to 0$ or when $v_{\xi} \gg v$. 
 
\subsection*{Stueckelberg mechanism for $\boldsymbol{Z_{\text{BL}}}$} 
\label{subsec:stue}

The generation of gauge boson masses takes place via two mechanisms.
For the gauge fields associated with the SM gauge group, masses are generated via the Higgs mechanism, which is triggered when $H$ gets a vev.
On the other hand, since $U(1)_{B-L}$ remains exact, $Z_{\text{BL}}$ -- the associated gauge field -- becomes massive via the Stueckelberg mechanism, which we summarise in what follows.

We start by writing down the kinetic Lagrangian\footnote{
We neglect tree-level kinetic mixing. 
This will be generated at one loop level, by the particles charged under $U(1)_{Y}$ and $U(1)_{B-L}$ \cite{Gherghetta:2019coi}.
In our benchmarks the loop-induced kinetic mixing parameter is small $\epsilon\lsim\mathcal{O}(10^{-3})$, evading the general constraints discussed in \cite{Williams:2011qb}.
}
for $Z_{\text{BL}}$ \cite{Ruegg:2003ps}
  \begin{equation}\label{StuKin}
\mathcal{L}^{\mathrm{Stu}}_{\mathrm{kin}}=-\frac{1}{4}Z_{\text{BL}}^{\mu\nu}Z_{BL\mu\nu}+\frac{1}{2}(m_{\text{BL}}Z_{\text{BL}}^{\mu}-\partial^\mu A)^2,
\end{equation}
where $Z_{\text{BL}}^{\mu\nu}=\partial^\mu Z_{\text{BL}}^\nu-\partial^\nu Z_{\text{BL}}^{\mu}$ and $A$ is the Stueckelberg scalar. 
In order for Eq. (\ref{StuKin}) to be gauge invariant, not only $Z_{\text{BL}}$ but also $A$ needs to transform under $U(1)_{B-L}$ as
\begin{equation}
\begin{split}
Z_{\text{BL}}^{\mu}& \to Z_{\text{BL}}^{\mu}+\partial^\mu\Lambda,\\
A& \to A+ M_{Z'}\Lambda.
\end{split}
\end{equation}
Next, we add to Eq. (\ref{StuKin}) the gauge-fixing term below
\begin{equation}
\mathcal{L}^{\mathrm{Stu}}_{\mathrm{gf}}=-\frac{1}{2\omega}(\partial_{\mu} Z_{\text{BL}}^{\mu}+m_{\text{BL}} \omega A)^2,
\end{equation}
and find, up to a total derivative,
\begin{equation}
\mathcal{L}^{\mathrm{Stu}}_{\mathrm{kin}}+\mathcal{L}^{\mathrm{Stu}}_{\mathrm{gf}}=-\frac{1}{4}Z_{\text{BL}}^{\mu\nu}Z_{BL\mu\nu}+\frac{1}{2}m^2_{BL}Z_{\text{BL}}^{\mu}Z_{BL\mu}-\frac{1}{2\omega}(\partial_{\mu} Z_{\text{BL}}^{\mu})^2+\frac{1}{2}\partial^\mu A \partial_\mu A -\frac{1}{2}m^2_{BL} \omega A^2,
\end{equation}
from which we can easily see that the Stueckelberg field decouples and $Z_{\text{BL}}$ gets a gauge invariant mass $m_{\text{BL}}$.
The latter, in contrast to gauge boson masses generated via the Higgs mechanism, does not depend on any vev or gauge coupling.

Finally, for the sake of completeness, we provide the relevant $Z_{\text{BL}}$ interaction terms
\begin{eqnarray}
\mathcal{L}^{Z_{\text{BL}}}_{\text{f}} &=& g_{\text{BL}}~Z_{BL\mu}\sum_{i=1}^{3}\left[\frac{1}{3}(\overline{u}_i\gamma^\mu u_i+\overline{d}_i\gamma^\mu d_i)-\overline{e}_i\gamma^\mu e_i-\overline{\nu}_i\gamma^\mu \nu_i+2\overline{S}\gamma^\mu S\right],\\
\mathcal{L}^{Z_{\text{BL}}}_{s}&=&3 i g_{\text{BL}}~Z_{BL\mu}\left[ \eta^{-}\partial^\mu \eta^{+}-\eta^{+} \partial^\mu\eta^{-} +\sum_{i=1}^{2}\left(\varphi^{0*}_i\partial^\mu \varphi^{0}_i- \varphi^{0}_i\partial^\mu\varphi^{0*}_i \right)\right] \nn\\&&+ 9 g_{\text{BL}}^{2}~{Z}_{BL}^{\mu}{Z}_{BL\mu}\left( \eta^{-}\eta^{+} + \sum_{i=1}^{2}\varphi^{0*}_i\varphi^{0}_i\right), \\
\mathcal{L}^{Z_{\text{BL}}}_{\text{g}-\text{s}}&=& 6 e g_{\text{BL}}~Z_{\text{BL}}^{\mu }\left\{ \left[A_\mu+ \cot (2\theta_W) Z_\mu\right]\eta^{-}\eta^{+} -\csc (2\theta_W) Z_\mu \left|\varphi^{0}_1 \cos \theta_{\varphi} -\varphi^{0}_2 \sin \theta_{\varphi} \right|^2\right.\nn\\
&&\left.+  \frac{\csc \theta_W}{\sqrt{2}} \left[W^+_\mu\eta^-(\varphi^{0}_1 \cos \theta_{\varphi} -\varphi^{0}_2 \sin \theta_{\varphi})+\mathrm{ h.c.}\right]\right\}~,\label{eq:ZpInt}
\end{eqnarray}
where $\theta_{W}$ is the electroweak angle.

\section{Constraints}
\label{sec:constraints}

In scotogenic schemes dark matter may either be fermionic or scalar.
In our construction there are two possible dark matter candidates, namely the complex scalar $\varphi_{1}^{0}$ and the Dirac fermion $S_{1}$.
Given their production mechanism and the processes through which they furnish the relic abundance, both of these candidates are WIMP-like. 
First of all, WIMP dark matter is subject to the observational bound on the cold DM relic\footnote{Smaller relic abundance would be allowed, however, in the presence of extra dark matter candidates, such as an axion. } in Eq.~(\ref{eq:omega}).
Measurement of nuclei recoils induced by the scattering of the local dark matter wind provides a direct WIMP detection/discovery method~\cite{Profumo:2019ujg}. 
The most recent limit for spin-independent DM-nucleon cross section is set by the Xenon1T collaboration \cite{Aprile:2018dbl}. 

Other phenomenological limits faced by our setup are summarized below in order to ensure that the parameter space within which we perform our numerical dark matter analysis is phenomenologically consistent. 

\subsection{Collider constraints}
\label{subsec:invisible}

\noindent
\underline{\textit{Dilepton searches:}}
Dilepton events would be induced at LEP and also at the LHC through the Drell-Yan mechanism.
Dilepton final state searches at these experiments with $36.1\text{ fb}^{-1}$ luminosity~\cite{Heeck:2014zfa} rule out values of $m_{\text{BL}}/g_{\text{BL}}$ for a new $Z_{\text{BL}}$ not satisfying the following condition
\begin{equation}
m_{\text{BL}}/g_{\text{BL}}\geq 6.9\text{ TeV}
\end{equation}
at 95\% C.L. 

\noindent
\underline{\textit{Invisible Higgs decay:}}
It is well-known that theories with continuous global symmetries spontaneously broken at accessible scales $\lsim \rm{few}~TeV$ lead to Goldstone bosons that can
couple to the Higgs. 
An example is the invisible Higgs decay to the invisible Majorons~\cite{Joshipura:1992hp,Bonilla:2015uwa,Bonilla:2015jdf,Fontes:2019uld}. 
In the present model the role of the Majoron is played by the Diracon. 
The branching ratio for $H_{1}$ decaying into a pair of Diracons is given by
\begin{equation}
  \label{eq:inv}
\text{Br}\bigl( H_{1}\to \mathcal{D}\mathcal{D} \bigr)=\Gamma\bigl( H_{1}\to \mathcal{D}\mathcal{D} \bigr)\cdot(\Gamma_{H_{1}}^{\text{SM}}+\Gamma\bigl( H_{1}\to \mathcal{D}\mathcal{D} \bigr))^{-1}.
\end{equation}
To estimate this branching, we use Eq.~(\ref{eq:V}) to write the Diracon couplings to the CP- and matter-parity-even scalars as
\begin{equation}
g_{H_{1}\mathcal{D}\mathcal{D}}=\dfrac{\lambda_{H\xi}}{2}v\cos{\theta_{h}}+\lambda_{\xi}v_{\xi}\sin{\theta_{h}}=\dfrac{m_{H_{1}}^{2}\sin{\theta_{h}}}{2v_{\xi}},~~~~~g_{H_{2}\mathcal{D}\mathcal{D}}=-\dfrac{\lambda_{H\xi}}{2}v\sin{\theta_{h}}+\lambda_{\xi}v_{\xi}\cos{\theta_{h}}=\dfrac{m_{H_{2}}^{2}\cos{\theta_{h}}}{2v_{\xi}}~. \label{eq:gHkDD}
\end{equation}
In the second equalities above we have expressed the quartic couplings $\lambda_{H\xi}$, $\lambda_{\xi}$ in terms of the squared mass splitting $\Delta m^{2}\equiv (m_{H_{2}}^{2}-m_{H_{1}}^{2})$ and mixing angle
\begin{equation}
\lambda_{H\xi}=\dfrac{-\Delta m^{2}\sin(2\theta_{h})}{2vv_{\xi}},~~~~~\lambda_{\xi}=\dfrac{1}{2v_{\xi}^{2}}\biggl( m_{H_{1}}^{2}\sin^{2}{\theta_{h}}+(m_{H_{1}}^{2}+\Delta m^{2})\cos^{2}{\theta_{h}} \biggr)~.
\label{eq:tradeQuartics}
\end{equation}
Note that the Higgs invisible decay width to Diracons~\cite{Bonilla:2016zef,Bonilla:2016diq},  
\begin{equation}
\Gamma\bigl( H_{1}\to \mathcal{D}\mathcal{D} \bigr)=\dfrac{m_{H_{1}}^{3}\sin^{2}{\theta_{h}}}{32\pi v_{\xi}^{2}}~, \label{eq:BRtoDD}
\end{equation}
only depends on $\sin{\theta_{h}}$ and $v_{\xi}$ and is suppressed by the Higgses mixing angle.
This is subject to the bound on Higgs decay to invisible states~\cite{Aaboud:2019rtt,Sirunyan:2018owy}
\begin{equation}
  \label{eq:inv-limit}
\text{Br}\bigl( H_{1}\to \text{invisible} \bigr)\lesssim 0.24 ,
\end{equation}
which is shown as a gray band in the left panel of Fig. \ref{fig:Neff} along with cosmological constraints, which are addressed in the next subsection.

\subsection{Effective number of neutrinos and cosmology}
\label{subsec:neff}

The presence of extra light degrees of freedom can alter the Hubble expansion in the radiation dominated epoch and therefore set stringent constraints from Big Bang nucleosynthesis
(BBN)~\cite{GonzalezGarcia:1989py} and CMB~\cite{Aghanim:2018eyx} observations.
In many $Z'$ setups the light states subject to these constraints are the right-handed neutrinos, whose thermalization with the SM can be mediated by the $B-L$ gauge boson if it appears sufficiently coupled in the effective theory. 
In our scenario, in addition to $\nu_{R}$, the massless $\mathcal{D}$ are also subject to these constraints.

\begin{figure}[h!]
\centering
\includegraphics[scale=0.55]{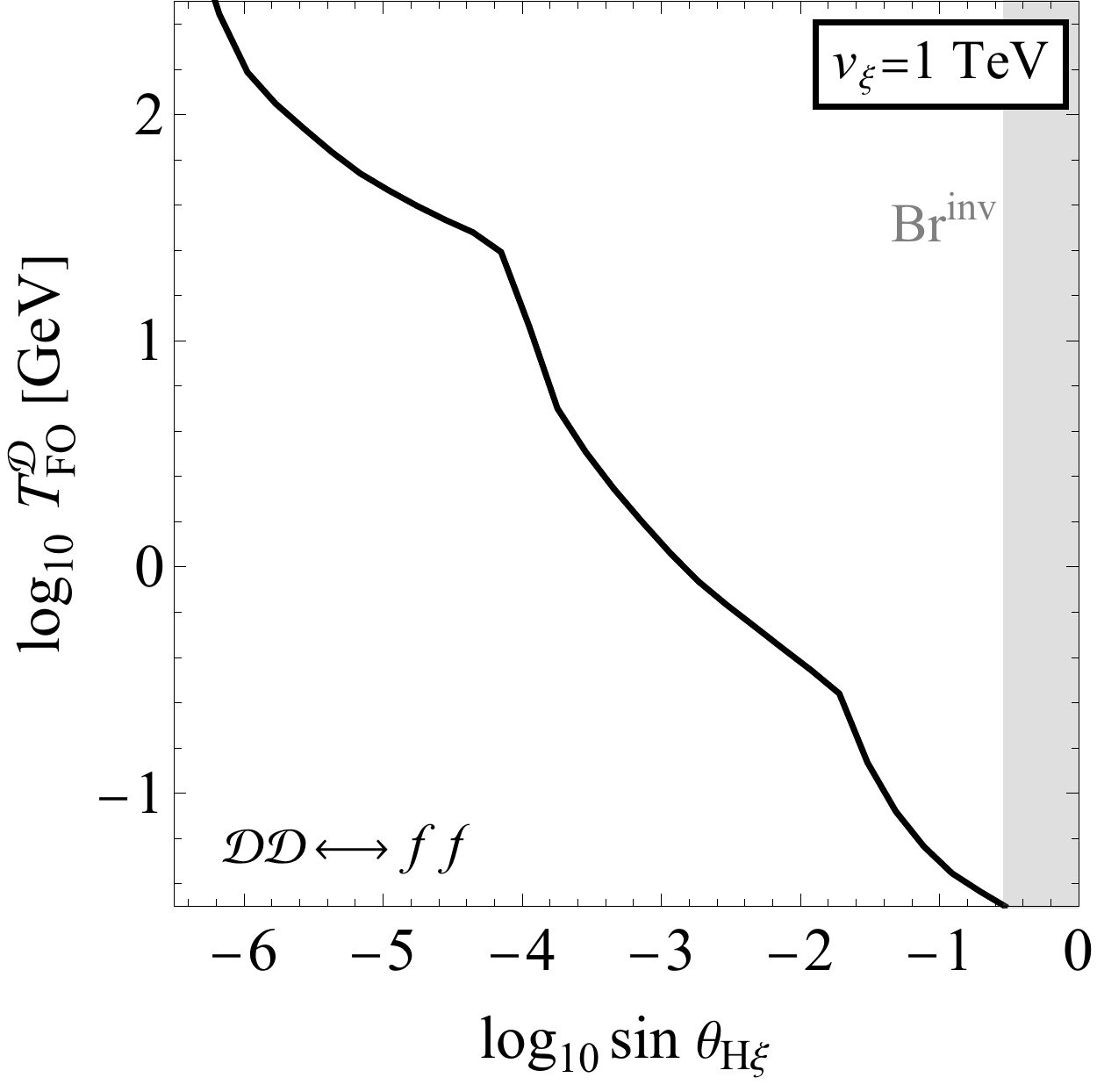}
\hspace{5mm}
\includegraphics[scale=0.55]{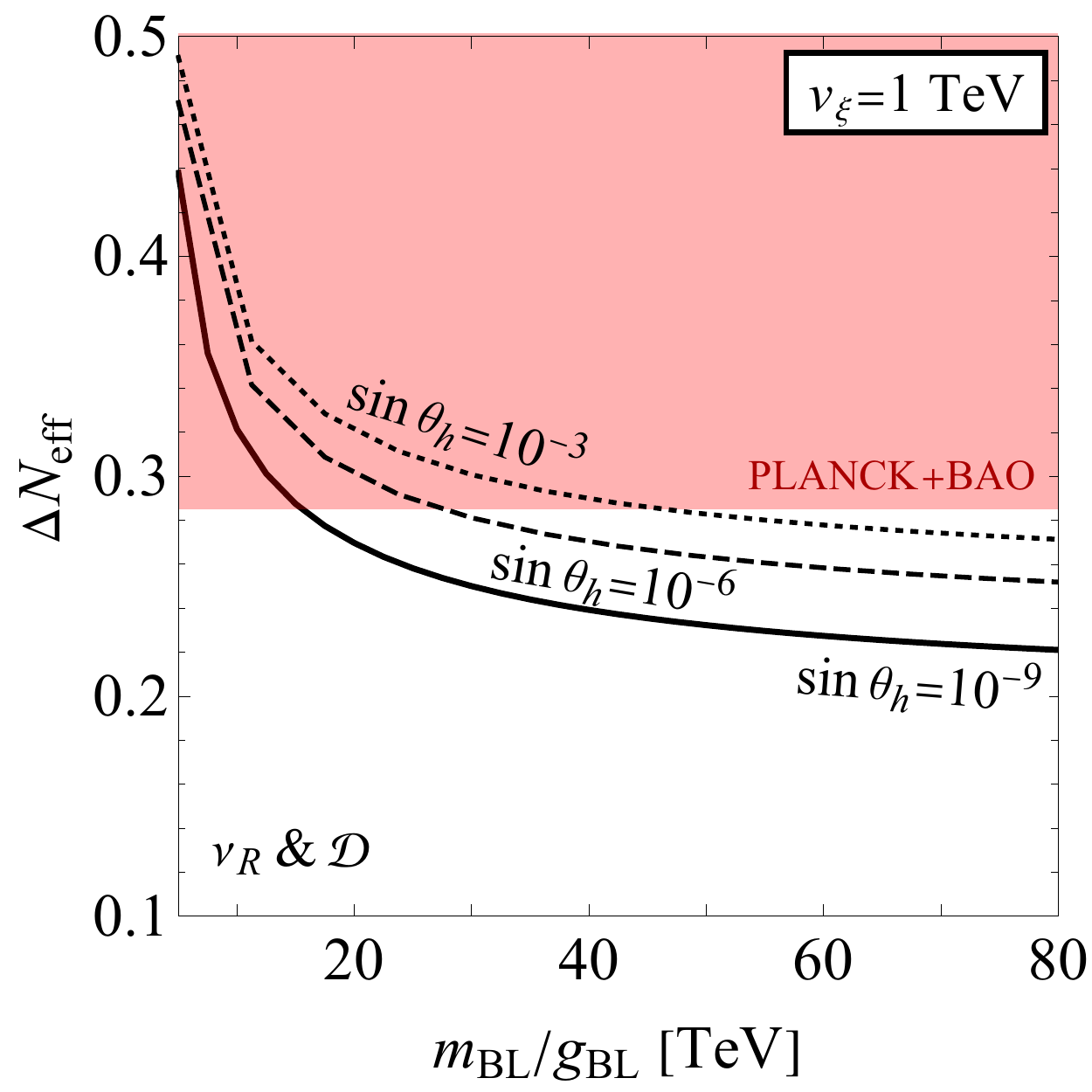}
\caption{ \textbf{Left:} Decoupling temperature of $\mathcal{D}\mathcal{D}\leftrightarrow f\bar{f}$ as a function of $\sin{\theta_{h}}$.
  Invisible Higgs decay bound is shaded gray.
  \textbf{Right:} Planck+BAO constraint on $\Delta N_{\text{eff}}$ from the combined contributions from $\nu_{R}$ and $\mathcal{D}$, for $T_{\text{FO}}^{\mathcal{D}}\gg T_{\text{FO}}^{\nu_{R}}$
  and at several $\sin{\theta_{h}}$.      
} \label{fig:Neff}
\end{figure}

While other $\nu_{R}$ and $\mathcal{D}$ decoupling situations may be possible, here we assume that the Goldstone coupling is weaker than the $B-L$ gauge interactions of right-handed neutrinos (RHNs).
Indeed, for a very tiny Higgs mixing angle $\theta_h$, one can have the Diracon freeze-out temperature around $T_{FO}^{\mathcal{D}}\sim O(100\text{ GeV}-1\text{ TeV})$, shown in Fig. \ref{fig:Neff}. 
This range of Diracon decoupling temperatures simplifies the counting of relevant degrees of freedom (DOFs), and the evolution of the radiation bath goes as follows.
As long as the states $\varphi_{1,2},\eta^{+},S_{k},Z_{\text{BL}}$ are non-relativistic at Diracon decoupling (i.e. their masses are above $T_{\text{FO}}^{\mathcal{D}}$), the radiation bath is composed of the relativistic states, in this case $\text{SM}+\mathcal{D}+\nu_{R}$.
Below $T_{FO}^{\mathcal{D}}$, the radiation left is $\text{SM}+\nu_{R}$, with the decoupling of RHNs happening until $T_{FO}^{\nu_{R}}$. The approximate Hubble rate in this temperature range becomes 
\begin{equation}
\dfrac{H(T)}{\sqrt{\pi^{2}/90}}\approx \begin{cases}
(g_{*}^{\text{SM}}(T)+\tfrac{7}{8}\cdot3g_{\nu_{R}}+g_{\mathcal{D}})^{1/2}\dfrac{T^{2}}{M_{P}}, & T\gtrsim T_{FO}^{\mathcal{D}} \\
(g_{*}^{\text{SM}}(T)+\tfrac{7}{8}\cdot3g_{\nu_{R}})^{1/2}\dfrac{T^{2}}{M_{P}}, & T_{FO}^{\nu_{R}}<T<T_{FO}^{\mathcal{D}} \\
g_{*}^{\text{SM}}(T)^{1/2}\dfrac{T^{2}}{M_{P}}, & T<T_{FO}^{\nu_{R}}~.
\end{cases} \label{eq:Hubble}
\end{equation}
where $g_{*}^{\text{SM}}(T)$ is the effective number of SM degrees of freedom at temperature $T$, and $g_{\nu_{R}}=2$, $g_{\mathcal{D}}=1$ are the spin degrees of freedom of the RHN and Diracon, respectively.

We now obtain the $\nu_{R}$ and $\mathcal{D}$ interaction rates to compare them with the cosmological expansion rate $H(T)$. 
The $\nu_{R}$ keep thermal contact with the SM through the reaction $\nu_{R}\overline{\nu}_{R}\to Z_{\text{BL}}\to f\overline{f}$ with $f=q,\ell$. 
Hence $T_{FO}^{\nu_{R}}$ is calculated in the instantaneous freeze-out approximation via
\begin{equation}
n_{\nu_{R}}(T_{FO}^{\nu_{R}})\bigl\langle v\sigma[\nu_{R}\overline{\nu}_{R}\to Z_{\text{BL}}\to f\overline{f}] \bigr\rangle\approx H(T_{FO}^{\nu_{R}})~. \label{eq:instantaneousFO}
\end{equation}
The RHN number density is $n_{\nu_{R}}(T)\approx (3/4)(g_{\nu_{R}}\zeta(3)/\pi^{2})T^{3}$.
In the limit $s,T^{2}\lesssim m_{\text{BL}}^{2}$ the cross section in the above thermal-average $\vev{v\sigma}$ simplifies to \cite{SolagurenBeascoa:2012cz,FileviezPerez:2019cyn}  
\begin{equation}
\sigma(s)\approx \dfrac{Q_{\nu_{R}}^{2}}{12\pi}\left(\dfrac{g_{\text{BL}}}{m_{\text{BL}}}\right)^{4}\sum_{f}N_{f}^{C}Q_{f}^{2}(s+2m_{f}^{2})\sqrt{1-4m_{f}^{2}/s}~
\end{equation}
where $N_{f}^{C}$ and $Q_f$ are the number of colours and the $B-L$ charge of the fermion $f$, respectively.

It is important to keep the $m_{f}$ terms in $\sigma(s)$ above in order to correctly keep track of the number of thermalized fermions at each mass threshold. 
The thermal averaged cross section is evaluated with standard methods~\cite{GONDOLO1991145} and scales as $\propto T^{5}(m_{\text{BL}}/g_{\text{BL}})^{-4}$, and $T_{\text{FO}}^{\nu_{R}}$ is given from Eq. (\ref{eq:instantaneousFO}). 

For the Diracon the most important process contributing to $\mathcal{D}\mathcal{D}\to\text{SM}$ is the scattering with light fermions via $s$-channel, $\mathcal{D}\mathcal{D}\to H_{k}\to ff$.
The corresponding cross section in the low $T$ limit, i.e. for $s\lesssim m_{H_{k}}^{2}$, reads
\begin{equation*}
\sigma(s)\approx \dfrac{1}{2\pi}\sum_{f}N_{f}^{C}\left[ 1-\dfrac{4m_{f}^{2}}{s} \right]^{3/2}\left\{ \dfrac{g_{H_{1}\mathcal{D}\mathcal{D}}^{2}\kappa_{H_{1}ff}^{2}}{m_{H_{1}}^{4}} + \dfrac{g_{H_{2}\mathcal{D}\mathcal{D}}^{2}\kappa_{H_{2}ff}^{2}}{m_{H_{2}}^{4}} + 2\dfrac{g_{H_{1}\mathcal{D}\mathcal{D}}~\kappa_{H_{1}ff}~g_{H_{2}\mathcal{D}\mathcal{D}}~\kappa_{H_{1}ff}}{m_{H_{1}}^{2}m_{H_{2}}^{2}} \right\}
\end{equation*}
where dimensionful $g_{H_{1}\mathcal{D}\mathcal{D}}$ and $g_{H_{2}\mathcal{D}\mathcal{D}}$ are in Eq.~(\ref{eq:gHkDD}) and the dimensionless $\kappa_{H_{1}ff}$, $\kappa_{H_{2}ff}$ follow from Eq.~(\ref{eq:Yuk}),
\begin{equation}
\kappa_{H_{1}ff}=-(m_{f}/v)\cos{\theta_{h}}~,~~~~~\kappa_{H_{2}ff}=(m_{f}/v)\sin{\theta_{h}}~.
\end{equation}
  %
  Notice that the $H_{1}$ coupling to $\mathcal{D}$ is suppressed at small $\sin{\theta_{h}}$ but its Yukawa to SM fermions is not, and vice versa  for $H_{2}$. 
  Despite the apparent suppression of the $H_{2}$ piece by its propagator, this is canceled with the Higgs-Diracon couplings Eq.~(\ref{eq:gHkDD}), proportional to the squared mass.
  Hence $H_{2}$-exchange cannot be neglected.

  The $\vev{ v\sigma}$ is used in conjunction with (relativistic) Diracon number density $n_{D}(T)=g_{D}(\zeta(3)/\pi^{2})T^{3}$ to get the Diracon freeze-out temperature.
  In the left panel of Fig. \ref{fig:Neff} we give $T_{\text{FO}}^{\mathcal{D}}$ as a function of $\sin \theta_{h}$, together with the band which is currently ruled out by LHC.
  The latter comes from the invisible Higgs decay bound in Eq.~(\ref{eq:inv-limit}).

  One usually parametrizes the contribution of a dark radiation species $X$ in terms of the effective number of neutrinos $\Delta N_{\text{eff}}$.
  There are two contributions, coming from $X=\nu_{R}$ and $X=\mathcal{D}$, both of which decouple while relativistic.  
  For $X=\nu_{R}$ the $\Delta N_{\text{eff}}$ contribution can be expressed in terms of the $\nu_{R}$ radiation density and active neutrino ($\nu$) temperature as

\begin{equation}
\Delta N_{\text{eff}}\supset \dfrac{\rho_{\nu_{R}}(T_{\nu_{R}})}{
2\tfrac{7}{8}\tfrac{\pi^{2}}{30}T_{\nu}^{4}
}=N_{\nu_{R}}\left( \dfrac{11}{4} \right)^{4/3}\left[ \dfrac{g_{*\text{CMB}}^{s}}{g_{*}^{s}(T_{FO}^{\nu_{R}})} \right]^{4/3}~~~~~\text{at }T=T_{\text{CMB}} \label{eq:NeffnuR}
\end{equation}
which follows directly from conservation of the SM and $\nu_{R}$ entropy densities and their ratio \cite{Blennow:2012de}.

The above DOF ratio follows from the $T_{\nu_{R}}(t)/T_{\gamma}(t)$ ratio at CMB temperature. 
The case of $\mathcal{D}$ is slightly more involved, as the $T_{\mathcal{D}}(t)/T_{\gamma}(t)$ ratio needed in $\Delta N_{\text{eff}}$ is not maintained from $T_{FO}^{\mathcal{D}}$ all the way to $T_{\text{CMB}}$. 
Rather, it gets modified at intermediate times when the $\nu_{R}$ decouples from the SM. Keeping track of the entropy densities one obtains,
\begin{equation}
\Delta N_{\text{eff}}\supset \dfrac{\rho_{\mathcal{D}}(T_{\mathcal{D}})}{
2\tfrac{7}{8}\tfrac{\pi^{2}}{30}T_{\nu}^{4}
}=N_{\mathcal{D}}\dfrac{1}{2}\dfrac{8}{7}\left[ \dfrac{g_{*\text{CMB}}^{s}}{g_{*}^{s}(T_{FO}^{\mathcal{D}})} \right]^{4/3}\left[ 1+\dfrac{N_{\nu_{R}}\cdot\tfrac{7}{8}\cdot2}{g_{*}^{s}(T_{FO}^{\nu_{R}})} \right]^{4/3}~~~~~\text{at }T=T_{\text{CMB}}, \label{eq:NeffD}
\end{equation}
with $N_{\mathcal{D}}=1$ and $N_{\nu_{R}}=3$ denoting the number of species of the corresponding particles.
Notice that, the ratio in Eq.~(\ref{eq:NeffD}) is suppressed for larger $T_{FO}^{\mathcal{D}}$.  
Using the freeze-out temperatures $T_{FO}^{\mathcal{D}}$ and $T_{FO}^{\nu_{R}}$ we add the contributions in Eqs.~(\ref{eq:NeffnuR}) and (\ref{eq:NeffD}) to confront against the
limit set by~\footnote{Due to the Lithium abundance uncertainties the BBN limit on $\Delta N_{\text{eff}}$ has been superseded by that of the CMB.}
PLANCK+BAO \cite{Aghanim:2018eyx}
\begin{equation}
  \label{eq:neff}
N_{\text{eff}}=2.96^{+0.34}_{-0.33}  
\end{equation}
In the right panel of Fig. \ref{fig:Neff} one sees the resulting restrictions as a function of $m_{\text{BL}}/g_{\text{BL}}$ for various $\sin \theta_{h}$ values.  
One notices that the earlier the Diracon decouples (smaller $H-\xi$ mixing) the weaker the limits on $m_{\text{BL}}/g_{\text{BL}}$.

  For $\sin \theta_{h}\lesssim 10^{-9}$ (solid line) the $\mathcal{D}$ contribution is already negligible, so that the $\Delta N_{\text{eff}}$ curve is indistinguishable from that which results from RHNs only.
  Hence, when $\mathcal{D}$ does not contribute to dark radiation, $m_{\text{BL}}/g_{\text{BL}}\gtrsim 15$~TeV is the weakest possible bound.
  As expected, the Diracon effect becomes relevant for higher $H-\xi$ mixing, pushing the $m_{\text{BL}}/g_{\text{BL}}$ lower bound up by a few tens of TeV (dashed and dotted lines in Fig. \ref{fig:Neff}).

After reviewing the most important observational constraints on DM and/or the dark mediator(s), we can now numerically analyze the relevant parameter space of our model.

\section{ Dirac fermion scotogenic dark matter}
\label{sec:FermionDM}

In what follows we examine the phenomenology of the fermionic scotogenic dark matter scenario through a detailed numerical study.
We assume that the dark matter candidate is the singlet Dirac fermion $S_{1}$, while taking the other dark mediators running in the scotogenic loop heavier, so they can decay as $\varphi\to S_{1}\ell$ (from now on $\varphi=\varphi^{0}_{1,2},\eta^{\pm}$).
The only direct couplings of $S_{1}$ are its $B-L$ interactions and the Yukawa couplings in Eq.~(\ref{eq:Yuk}).
This implies that there are two generic \textit{portals} connecting our scotogenic WIMP dark matter to the SM particles.  
Each of these individual portals has been worked out extensively in the literature~\cite{Dong:2017zxo,Blanco:2019hah,Calle:2019mxn,Han:2020oet,Bai:2014osa,Okawa:2020jea,Blennow:2019fhy}. 
In order to comply with the LEP bounds on charged scalars, we restrict ourselves to dark scalar mediator masses $m_{\varphi_2,\eta^{+}}\gtrsim 100\text{ GeV}$~\cite{Heister:2002ev}\footnote{
  There are charged scalar mass limits from the LHC \cite{Aad:2019vnb,Aad:2019byo} which depend on assumptions concerning the dominant decay modes $\eta^\pm \to W^\pm \varphi^0_1$ and/or $\eta^\pm \to \ell^\pm S_1$ (with $\ell=e,\mu,\tau$) ~\cite{Farzan:2010fw,Okawa:2020jea}. Detailed analysis on this point is beyond the scope of our paper.}.
Notice that the $U(1)_{G}$ breaking scale $v_{\xi}$ controls both the $m_{\nu}$ loop size as well as the $H-\xi$ mixing parameter. In what follows we opt to fix a benchmark value $v_{\xi}=1$~TeV. 

\begin{figure}[h!]
\centering
\includegraphics[scale=0.55]{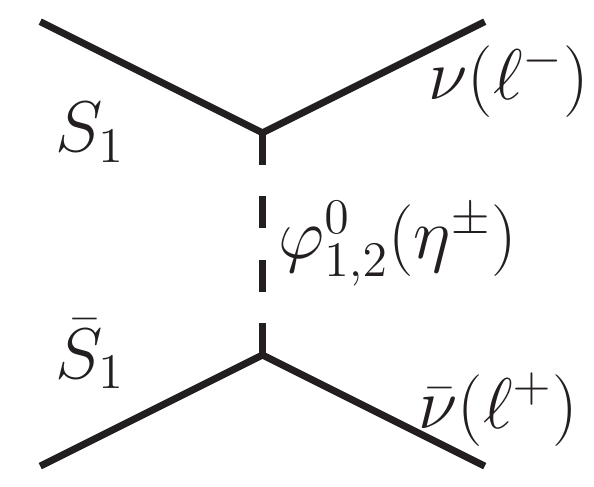}
\ \
\includegraphics[scale=0.55]{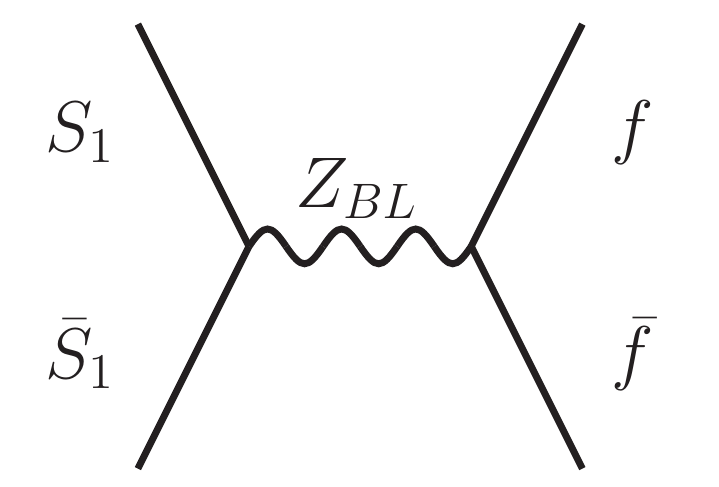}
\ \
\includegraphics[scale=0.55]{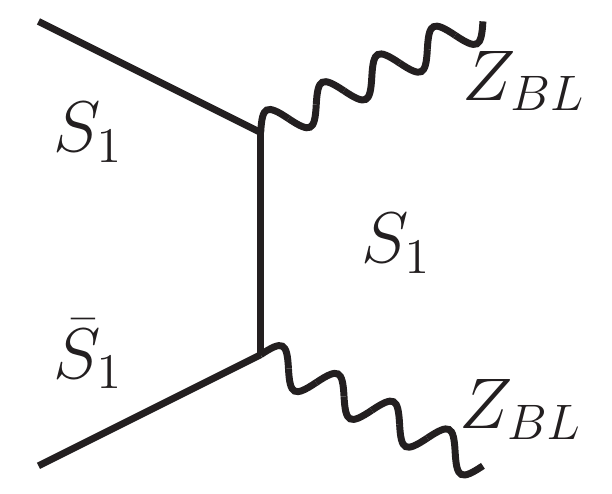}
\ \
\caption{DM pair annihilation modes, featuring $\varphi$-exchange (left) and $Z_{\text{BL}}$ portal (center and right).} 
\label{fig:AnnDiag}
\end{figure}

\begin{figure}[h!]
\centering
\includegraphics[scale=0.5]{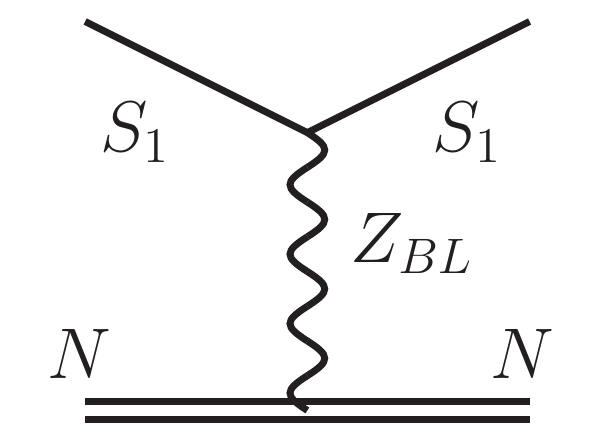}
\caption{Direct detection diagrams via the $Z_{\text{BL}}$ portal.}
\label{fig:DD}
\end{figure}

An important defining feature of scotogenic schemes is that dark matter candidates are also the mediators of neutrino mass generation \cite{Ma:2006km,Hirsch:2013ola,Merle:2016scw,Avila:2019hhv,Boehm:2006mi,Farzan:2009ji}.
As a result, restrictions from neutrino mass and dark matter phenomenology must be taken into account jointly, in order to characterize the relevant parameter space.
Given the correct $m_{\nu}$ scale, one should also account for the mass splittings and mixing angles observed in neutrino oscillations. 
As mentioned above, this requires the other dark mediators, e.g. fermions $S_{2},S_{3}$, which bring in the extra parameters. 
Relevant diagrams for $S_{1}$ annihilation are given in Fig. \ref{fig:AnnDiag}, while Fig. \ref{fig:DD} shows the dark matter scattering amplitude off nuclei.
One sees that dark matter scattering involves only the $Z_{\text{BL}}$-portal, while the dark scalars also take part in setting the relic density of dark matter.

\subsection{Scotogenic Dirac fermion DM}
\label{subsec:twoPortals} 

\begin{figure}[h!]
\includegraphics[scale=0.6]{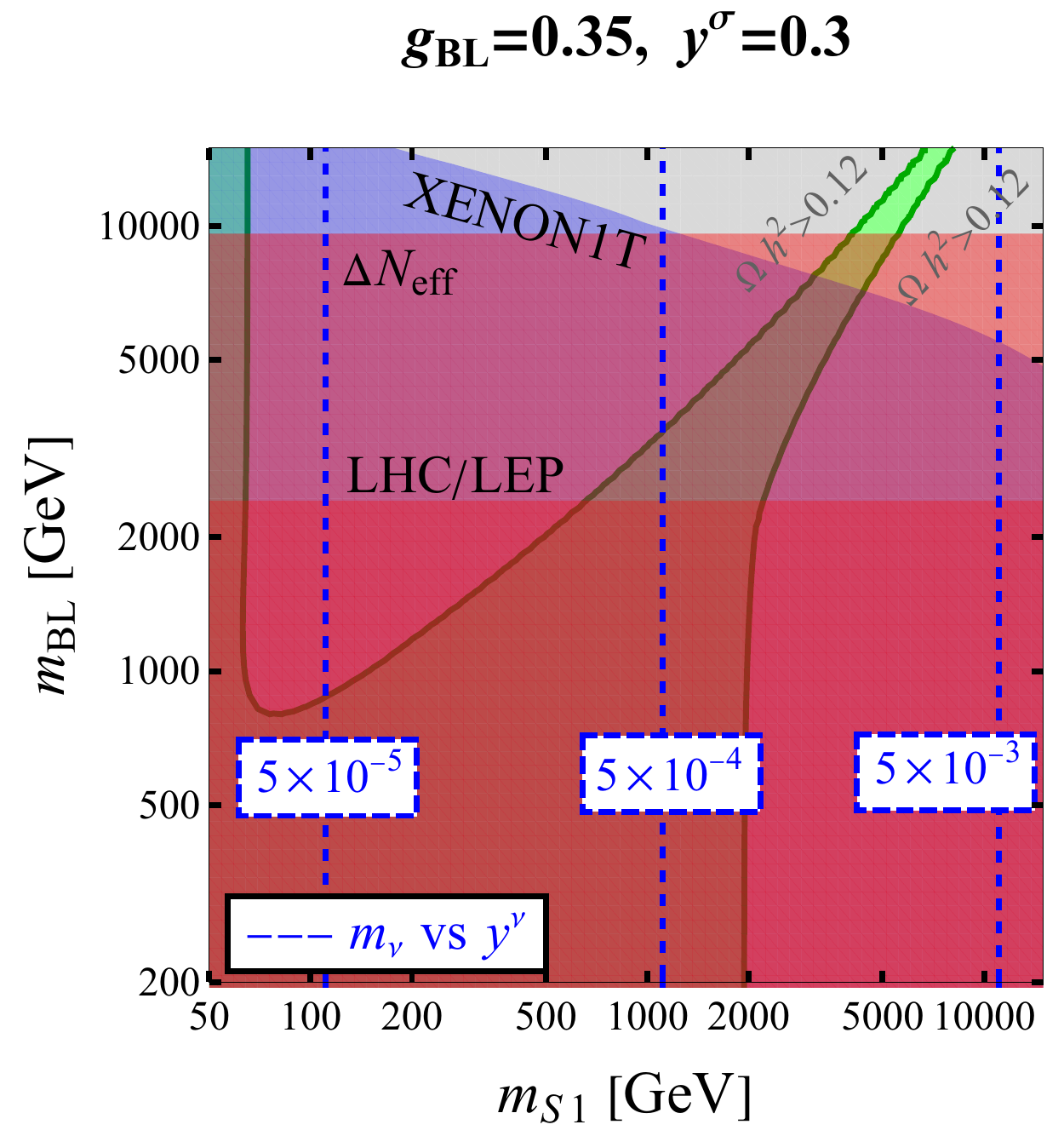}
\hspace{5mm}
\includegraphics[scale=0.6]{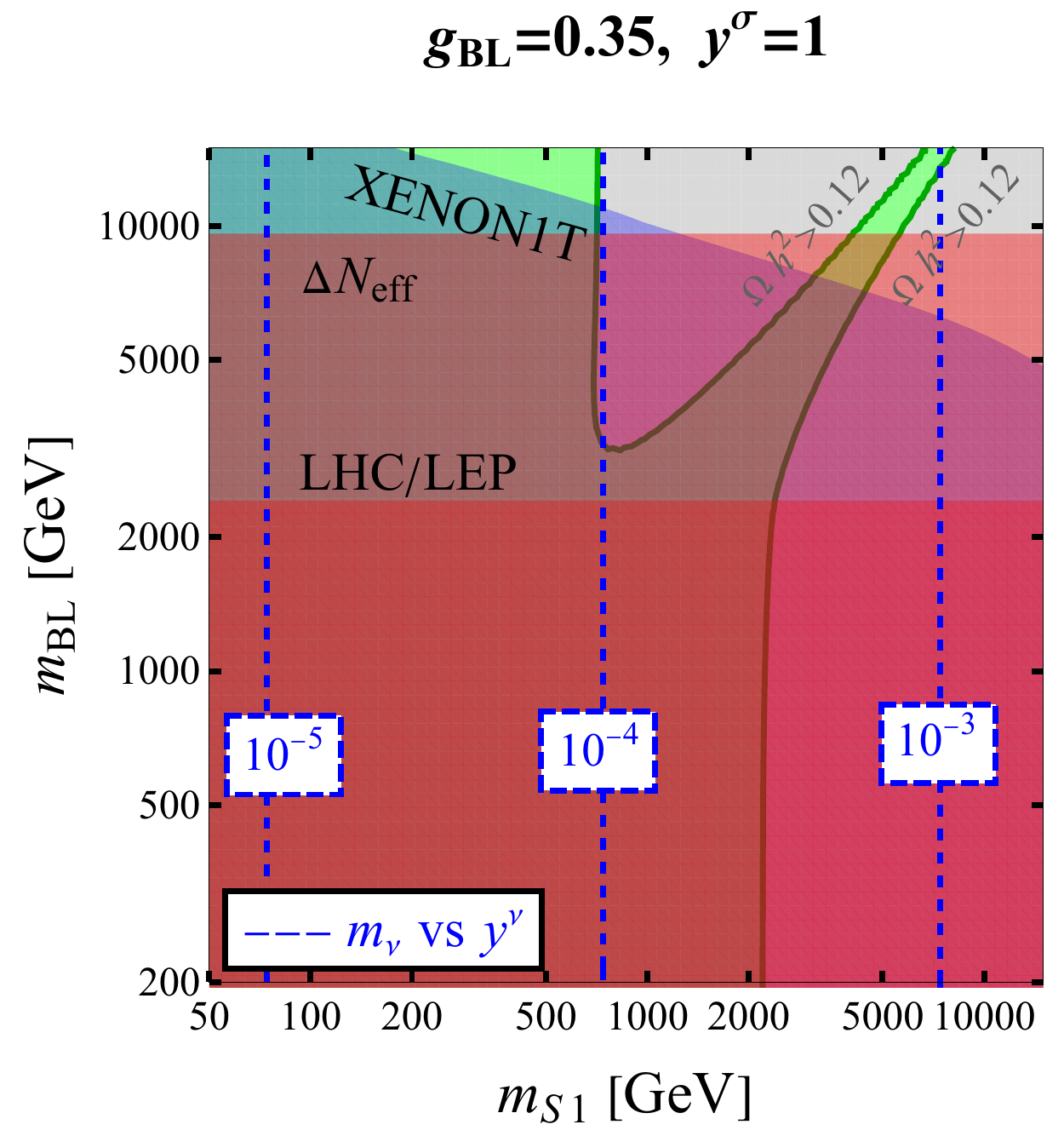}
\hspace{5mm}
\includegraphics[scale=0.6]{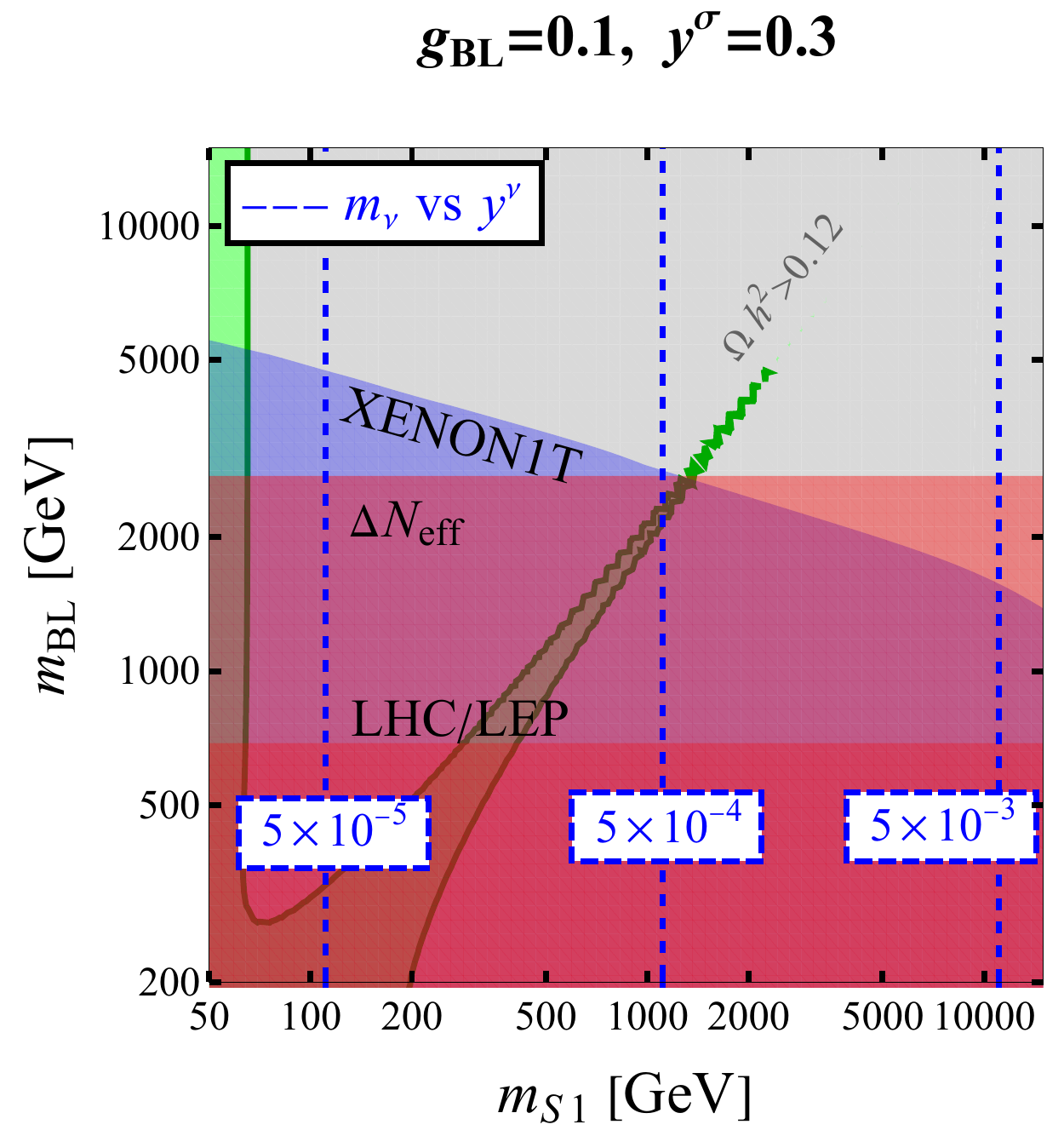}
\hspace{5mm}
\includegraphics[scale=0.6]{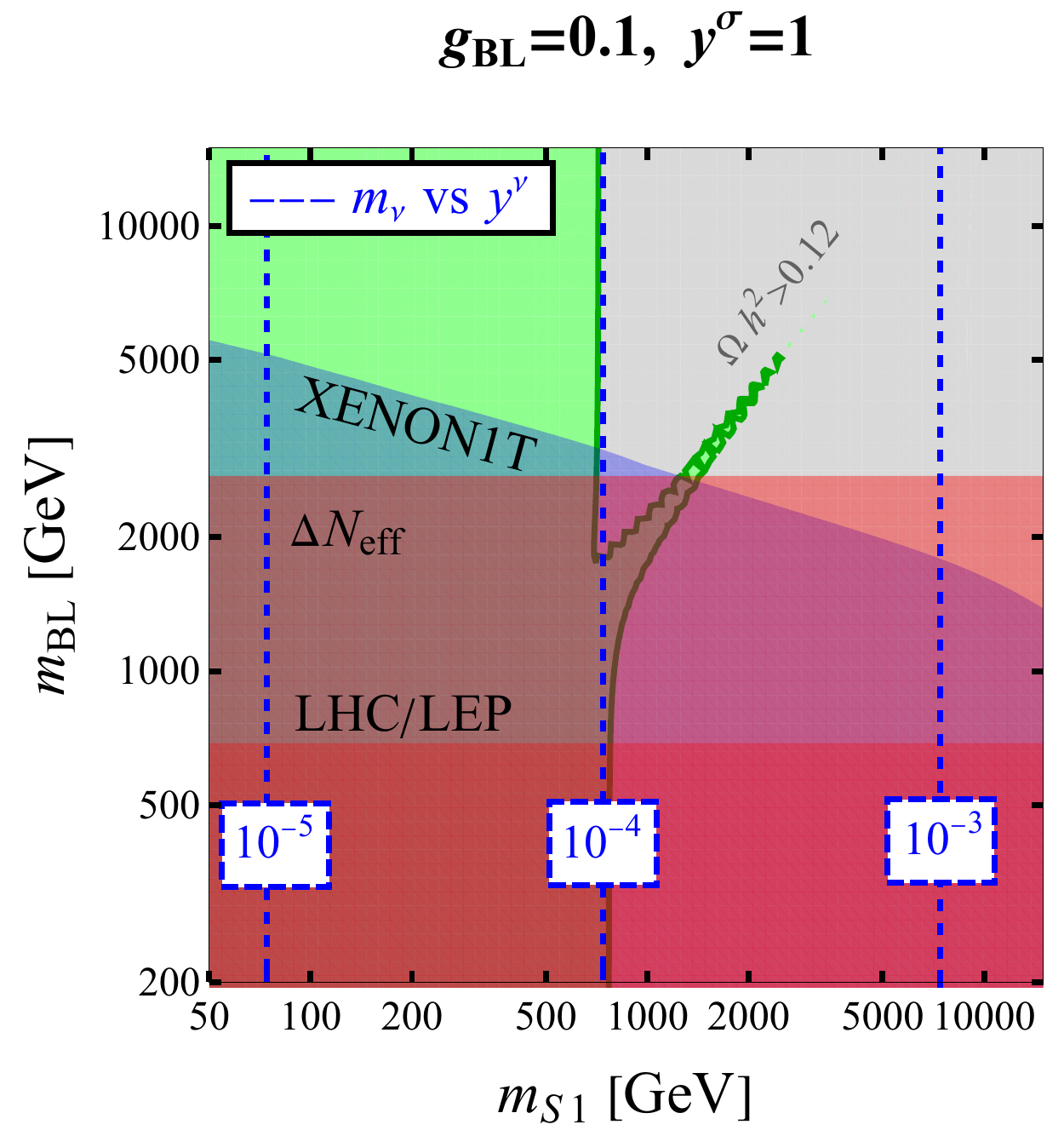}
\caption{Singlet Dirac fermion scotogenic dark matter  $S_{1}$ in the presence of $Z_{\text{BL}}$ and $\varphi$-exchange annihilation.
  Upper and lower panels have distinct gauge coupling $g_{\text{BL}}$, and left and right panels have distinct $S$-Yukawas.
  A $100\%$ DM relic occurs on the solid green line, and the regions excluded by LHC/LEP and $\Delta N_{\text{eff}}$ limits are shaded red.
  In all panels one has $\lambda_{D}v_{\xi}=10^{-3}\text{ GeV}$, $m_{\varphi1}/m_{S1}=1.1$, and $m_{\varphi2}/m_{\varphi1}=3$.
  Vertical dashed blue lines indicate the Yukawa coupling $y^{\nu}$ required to hit the atmospheric neutrino mass scale at the corresponding $m_{S_1}$.} \label{fig:ZBLNUportalPlots}
\end{figure}

To start with we notice that the DM phenomenology is determined by the following set of parameters 
$$\{m_{S_{1}},~g_{\text{BL}},~m_{\text{BL}}, m_{\varphi_{1}},~m_{\varphi_{2}},~ m_{\eta^\pm},~y^{\nu},~y^{\sigma},~\lambda_{D}v_{\xi}\}.$$
On the other hand, accommodating adequate magnitudes for the radiative $m_{\nu}$ involves all of these parameters except for $g_{\text{BL}},~m_{\text{BL}}$.
Dark matter annihilation rates and nucleon cross sections are computed using the \texttt{SARAH/SSP} spectrum generator \cite{Staub:2015kfa,Porod:SPHENO}, together with the
\texttt{MicrOmegas} code \cite{Belanger:2018ccd}.
In our numerical analysis we perform a scan on the $(m_{S_{1}},m_{\text{BL}})$ plane, at fixed $g_{\text{BL}}$, Yukawa couplings, and $\lambda_{D}v_{\xi}$, see Fig. \ref{fig:ZBLNUportalPlots}.
Other parameters that remain fixed can be read from Table \ref{tab:benchmarks}.  
\begin{table}[t]
	\begin{tabular}{|c|c|c|}
	\hline
	$Z_{\text{BL}}$-portal $+~\varphi\text{ exchange}$ &  $\lambda_{\alpha}$,~$\lambda^{(\prime)}_{H\alpha}$,~$\lambda_{\xi \alpha}\approx0$\ \ \text{(with $\alpha=\sigma,\eta$)}, ~~~~~$\lambda_{\xi}=0.1$,~~~~~$\lambda_{H\xi}=10^{-6}$, \\
	~ & $\lambda_{D}=10^{-6}$,~~~~~ $v_{\xi}=1\text{ TeV}$,~~~~~$\mu_{\eta}^{2}/\mu_{\sigma}^{2}=9$,~~~~~ $\mu_{\sigma}^{2}=(1.1~m_{S1})^{2}$ \\
	\hline
	\end{tabular}
	\caption{Model benchmarks.}
	\label{tab:benchmarks}
\end{table}
In the panels of Fig. \ref{fig:ZBLNUportalPlots} we illustrate the DM constraints in the general setup combining $Z_{\text{BL}}$ and $\varphi$-exchange annihilation for two
$g_{\text{BL}}$ and two $y^{\nu},y^{\sigma}$ choices.
The direct detection limit set by Xenon1T on the scattering cross section with nuclei is shaded blue. 
A correct DM abundance is obtained on the dark green line, with the light green (gray) region indicating under-abundance (over-abundance).  
The distinctive cusp-like, resonant feature of the pure $Z_{\text{BL}}$ portal is evident along $2m_{S_{1}}=m_{\text{BL}}$.  
On the other hand the lower region below the cusp feature corresponds to annihilation into $Z_{\text{BL}}$ pairs in the $t$-channel.
In the upper-left region of all panels, the vertical portion of the DM relic contour is due to $\varphi$-exchange. 
This channel makes this region allowed by Xenon1T at $m_{S_{1}}<70 (800)\text{ GeV}$ in the left (right) panels of Fig.~\ref{fig:ZBLNUportalPlots}, as long as $m_{\text{BL}}$ exceeds some minimal value.  
One sees how, for the chosen $g_{\text{BL}}$ values, the DM relic contour barely escapes the Xenon1T bound around the cusp (due to the $Z_{\text{BL}}$ portal) and the effect of the $\varphi$-exchange channel becomes visible.
Notice that this extra channel allows smaller $m_{S_{1}}$ values (light DM regime) compared to the pure $Z_{\text{BL}}$ channel.
%
%

Moreover, for our gauge coupling choices, the collider limits on $m_{\text{BL}}/g_{\text{\text{BL}}}$ (darker red shade) become weaker than those coming from direct DM detection and $\Delta N_{\text{eff}}$.
The latter limit dominates for dark matter masses above 1~TeV.  
In addition, the vertical blue dashed lines in Fig.~\ref{fig:ZBLNUportalPlots} indicate the particular DM mass $m_{S_{1}}$ corresponding to the atmospheric neutrino mass scale, 
given our Yukawa coupling choices~\footnote{This would correspond to the largest neutrino mass in a hierarchical normal-ordered neutrino mass spectrum.}. 
 Choosing larger $y^{\nu}$, $y^{\sigma}$ Yukawa couplings would shift the dark green vertical relic-density contour to the right, due to larger $\varphi$-exchange annihilation.
 However, this would affect $m_{\nu}$ more strongly, so that the $\sqrt{\Delta m_{\text{atm}}^{2}}$ benchmark would end up within the DM overabundant (gray) region. 
 Hence the combined requirements of DM abundance and generating the atmospheric neutrino mass \textit{scotogenically} pushes us to restricted dark matter masses and couplings
 that can be probed in nuclear recoil scattering. 
 In what follows we describe the special regimes where only one annihilation channel is available.

 \subsection{ $Z_{\text{BL}}$-portal limit}
\label{sub:ZBLOnly}

With $Z_{\text{BL}}$ present in the effective theory, singlet fermion DM annihilates as $S_{1}\overline{S}_{1}\to Z_{\text{BL}}\to f\overline{f}$ in the $s$-channel and $S_{1}\overline{S}_{1}\to Z_{\text{BL}}Z_{\text{BL}}$
in the $t$-channel (for $m_{\text{BL}}<m_{S_{1}}$ only)~\cite{Dong:2017zxo,Blanco:2019hah,Han:2020oet,Blennow:2019fhy}.
The final state of the former is mainly leptonic, due to the $B-L$ assignments, and electroweak bosons are absent in the final state for unmixed $Z_{\text{BL}}$.
Here the dark matter phenomena are determined by $\{m_{S_{1}},~m_{\text{BL}}$,~$g_\text{BL}\}$,
though it extends to the set $\{ m_{\varphi_{1}},~m_{\varphi_{2}}/m_{\varphi_{1}},~y^{\nu},~y^{\sigma},~\lambda_{D}v_{\xi} \}$ in order to accommodate the radiative $m_{\nu}$. 

Reaching a correct DM relic density relies mainly on the resonant $Z_{\text{BL}}$ annihilation near $2m_{S_{1}}=m_{\text{BL}}$, except when $Z_{\text{BL}}$ pair creation opens up at $m_{S_{1}}>m_{\text{BL}}$. 
Spin-independent scattering with nuclei occurs via $Z_{\text{BL}}$ exchange, and is expected to be important because the $Z_{\text{BL}}$ couples to quarks without suppression. 
Previous analyses have found that large regions in the $(m_{S_{1}},m_{\text{BL}})$ plane are excluded by over-abundance or direct detection bounds~\cite{Dong:2017zxo,Blanco:2019hah,Han:2020oet},
except for the tip of a cusp-like region along the $2m_{S_{1}}\approx m_{\text{BL}}$ resonant line. 

As mentioned in Sec. \ref{subsec:neff}, $Z_{\text{BL}}$ exchange between RHN and SM fermions contributes to $\Delta N_{\text{eff}}$, see solid curve in the right panel of Fig. \ref{fig:Neff}.
Notice that, once again assuming $\sin{\theta_h}\sim 10^{-6}$ -- in agreement with the benchmark in Table~\ref{tab:benchmarks} -- in such a way that the Diracon contributes to $\Delta N_{\text{eff}}$, the Planck+BAO limit becomes irrelevant for $m_{\text{BL}}/g_{\text{BL}} \gsim 30\text{ TeV}$.

\begin{figure}[h!]
\centering
\includegraphics[scale=0.6]{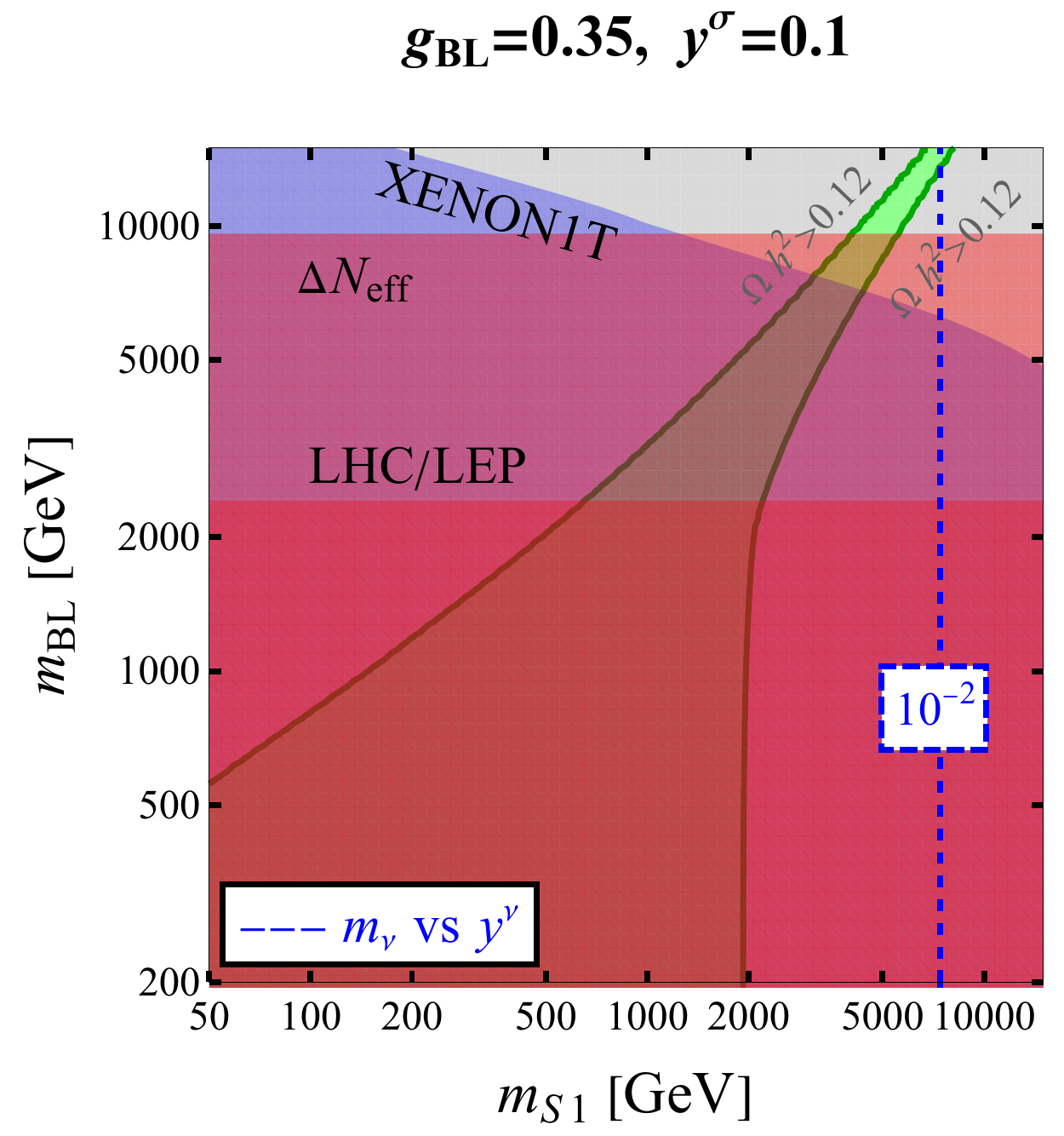}
\caption{Illustrating the case of singlet Dirac fermion scotogenic dark matter annihilating through $Z_{\text{BL}}$ only. As before,
100\% DM relic occurs on the solid green line, with color code as in Fig.~\ref{fig:ZBLNUportalPlots}. Notice that Xenon1T indicates a relatively large DM mass.
} \label{fig:ZBLlimit}
\end{figure}

\subsection{ $\boldsymbol{\varphi}$-exchange portal limit} 
\label{sub:phiOnly}
We now turn to the case where the DM annihilation is mediated by the dark scalars $\varphi$, i.e. only through the $t$-channel in Fig.~\ref{fig:AnnDiag}~\cite{Dong:2017zxo,Bai:2014osa,Okawa:2020jea,Blennow:2019fhy}.  
This means that the $Z_{\text{BL}}$ is sufficiently decoupled, either by taking $g_{\text{BL}}\ll~1$ or $m_{\text{BL}}\gg m_\text{DM}$.
In this case DM annihilation proceeds at tree-level via $S_{1}\overline{S}_{1}\to \nu\overline{\nu}$ and $S_{1}\overline{S}_{1}\to \ell^{-}\ell^{+}$, which respectively involve $\varphi_{1,2}^{0}$ or $\eta^{+}$ exchange, see left diagram in Fig. \ref{fig:AnnDiag}. 
There are also 1-loop channels mediated by SM gauge bosons, though these are found to be highly subleading for annihilation~\cite{Blennow:2019fhy}.
\begin{figure}[h!]
\centering
\includegraphics[scale=0.6]{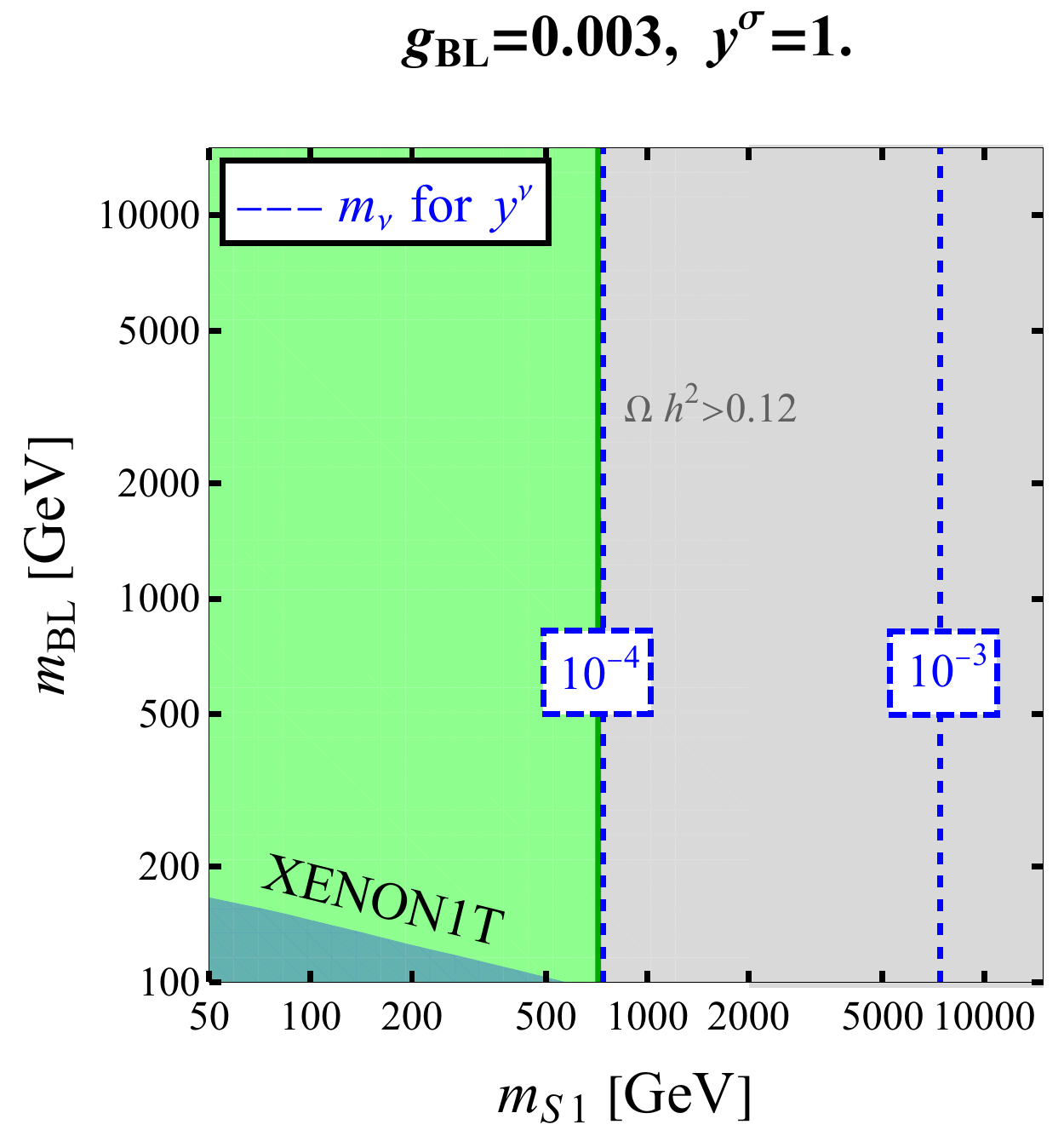}
\caption{Illustrating the case of singlet Dirac fermion scotogenic dark matter annihilating through $\varphi$-exchange only. As before 100\% DM relic occurs on the solid green line, adopting the same color code as in Fig.~\ref{fig:ZBLNUportalPlots}.
Notice that smaller DM mass values are consistent with Xenon1T results.
}
\label{fig:philimit}
\end{figure}

The parameters governing annihilation are the DM mass, the $\varphi$, $\eta^\pm$ mediator masses,  the Yukawa couplings of $S_{1}$, and the $\varphi$ mixing angle.  
This means the relevant parameter set is $\{ m_{S_{1}},~m_{\varphi_{2}},~m_{\varphi_{1}}, m_{\eta^\pm},~y^{\nu}, y^{\sigma},~\theta_{\varphi}\}$.
Note that DM annihilation channels mediated by $Z_{\text{BL}}$ can be suppressed by taking a smaller gauge coupling $g_{\text{BL}}$ value than those used in Fig.~\ref{fig:ZBLNUportalPlots}. 
The other parameters are fixed at the benchmarks given in Table \ref{tab:benchmarks}.
As expected, the direct detection cross section decreases as $g_{\text{BL}} \to 0$, indeed Fig.~\ref{fig:philimit} shows that the cusp-like green region disappears in this case. 

\begin{figure}[h!]
\centering
\includegraphics[scale=0.6]{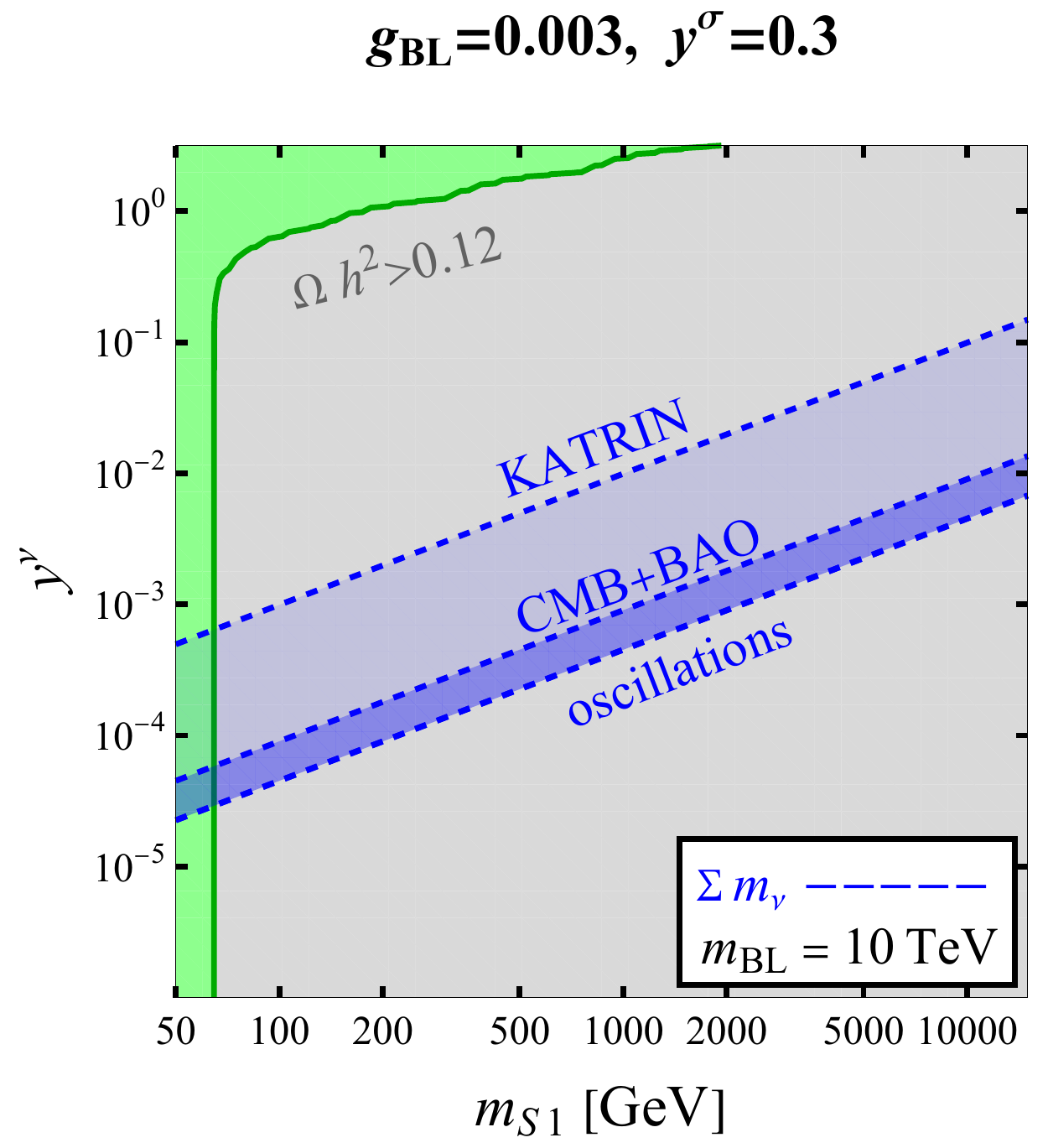}
\hspace{5mm}
\includegraphics[scale=0.6]{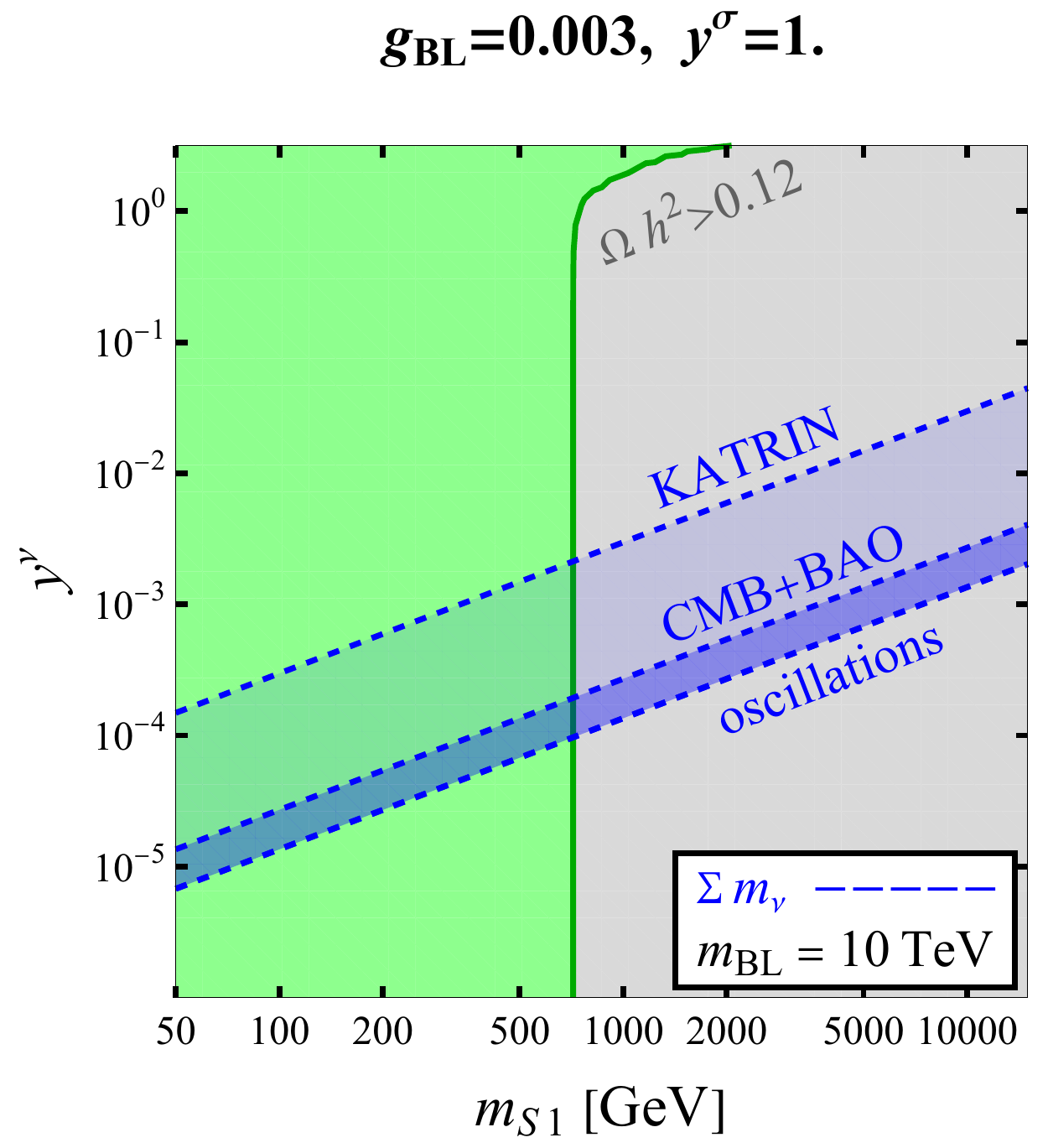}
\caption{Magnitude of the Dirac neutrino Yukawa coupling versus DM mass in the $\varphi$-exchange portal limit. The solid green line corresponds to $100\%$ DM relic.
  The light green(gray) region represents DM under(over)-abundance. The blue band is the region allowed by the requirement of a viable neutrino mass scale.} \label{fig:ynudm}
\end{figure}

Notice that, qualitatively, larger Yukawa couplings favor DM annihilation, but also increase the magnitude of the radiatively generated neutrino mass $m_{\nu}$. 
On the other hand, the radiative neutrino mass is also proportional to $\lambda_{D}$.
  One can reconcile having a viable $m_{\nu}$ and sufficient dark matter annihilation by a proper choice of $\lambda_{D}$ and the two Yukawa couplings.
  This is so because, for fixed $\lambda_{D}$, $m_{\nu}$ depends on the product of two Yukawas, while DM annihilation can be dominated by either of them separately.  
As a minimum estimate of the neutrino mass scale $m_{\nu}$~\footnote{Here we are not incorporating flavor symmetries, so we keep the discussion at the simplest, one-family level.}, we take Eq.~(\ref{eq:nuLoopMass}) neglecting flavor indices and set it to the \textit{atmospheric} mass splitting $\sqrt{\Delta m_{\text{atm}}^{2}}\approx 0.05\text{ eV}$~\cite{deSalas:2020pgw}.
    As a stringent limit we can take the cosmological one from Planck-2018~\cite{Aghanim:2018eyx},
    while a most conservative limit is to take the latest one from the KATRIN collaboration\footnote{In this case
  neutrinos would be nearly degenerate, for a detailed discussion see~\cite{Lattanzi:2020iik}. } ~\cite{Aker:2019uuj}~.
  Fig.~\ref{fig:ynudm} shows the consistency band obtained by requiring $m_{\nu}$ between 0.05~eV and 1 \text{eV}. One can then rule out a large chunk of the available space,
    i.e. the region below and above the band. Yet, since the estimate depends on other parameters such as $\lambda_{D}$ and the Yukawa couplings, this serves only for the sake of illustration.
    The intersection of the band with the relic density contour gives us an idea of the acceptable range of the Yukawa couplings.

It is worth mentioning that, in the $\varphi$-exchange regime, direct singlet dark matter detection does not place important constraints for masses $\lesssim 100\text{ GeV}$.
Indeed, scattering with nuclei only happens through loop diagrams involving $Z$ and Higgs boson exchange, and the loop diagrams are found subdominant~\cite{Blennow:2019fhy}. 
The relevant triangle and box diagrams become negligible in our model, since $m_{S_{1}}\gg m_{\nu}$.

\subsection{Low dark matter masses}
\label{sec:minimum-dark-matter}

As we saw above, see Fig.~\ref{fig:ynudm},  our scenario can accomodate light dark matter in the range where improved sensitivities in direct detection experiments are expected.
However, the Higgs boson does not decay to a pair of DM particles, as it is usually the case, if kinematically allowed. \\[-.2cm]
    
\begin{figure}[h!]
\centering
\includegraphics[scale=0.6]{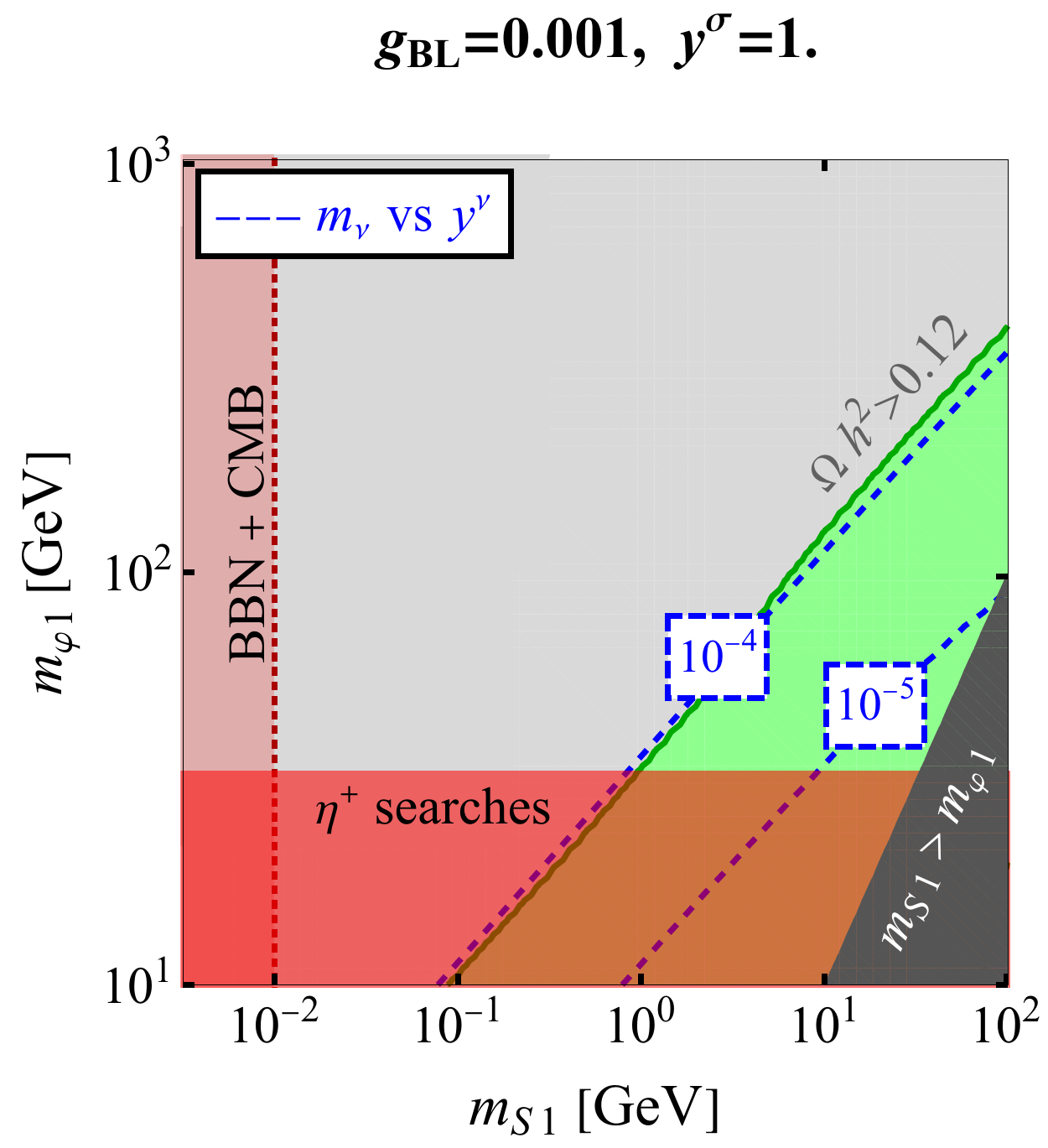}
\hspace{5mm}
\includegraphics[scale=0.6]{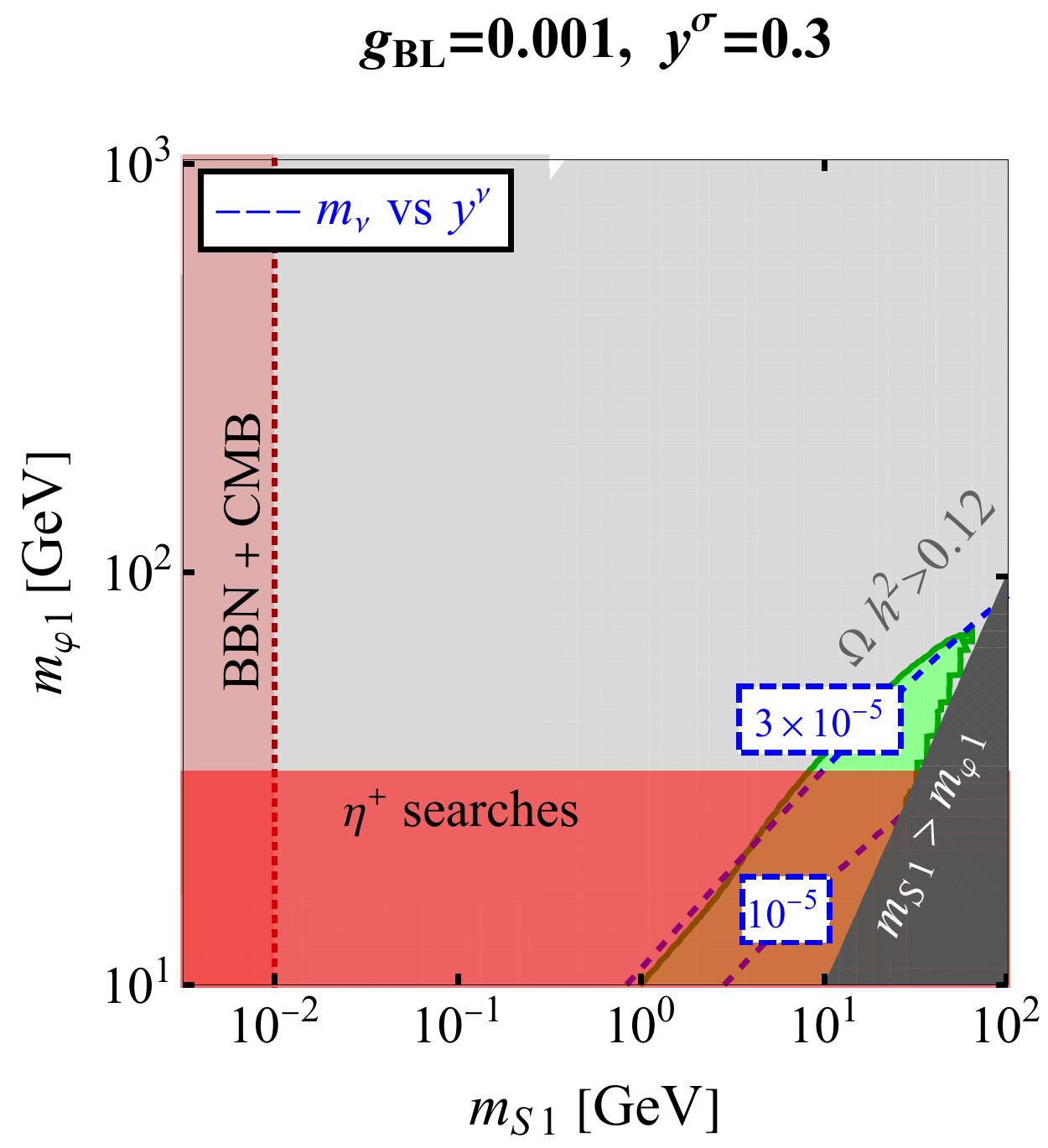}
\caption{ Smallest Dirac dark matter fermion mass within the $\varphi$-exchange annihilation scenario. In both panels we assume
$\lambda_{D}v_{\xi}=10^{-3}\text{ GeV}$, and the mediator masses satisfy $m_{\varphi2}/m_{\varphi1}=3$ and $m_{\eta^+}=m_{\varphi_2}\gtrsim 100$~GeV.}
\label{fig:FDMlight}
\end{figure}

Notice that our DM candidate could,  in principle, be much lighter. 
  However, there is a lower bound on the DM mass ($m_{S1}\gtrsim 10$~MeV) which is set by the cosmological constraints arising from primordial BBN~\cite{Iocco:2008va} and the CMB~\cite{Nollett:2014lwa}.
  In our setup we expect $m_{S1}\gtrsim1$~GeV  as illustrated in Fig.~\ref{fig:FDMlight}. In both panels the allowed parameter space region for light DM mass assumes $m_{\varphi2}/m_{\varphi1}=3$ and $m_{\eta^+}=m_{\varphi_2}\gtrsim 100$~GeV \cite{Heister:2002ev}.

As a final remark, we note, from Fig.~\ref{fig:ynudm}, that the allowed parameter region for DM and neutrino mass indicates $y^\nu\lesssim\mathcal{O}(10^{-3})$.
This Yukawa coupling of SM doublets also governs Lepton Flavor Violation (LFV), via processes such as $\mu\to e\gamma$.
With such Yukawa coupling sizes, the predicted LFV rates are in agreement with current experimental limits, see e.g.~\cite{Farzan:2012sa,Calle:2019mxn}.
Nonetheless, the Yukawa couplings of SM singlets, $y^\sigma$, are not constrained by these processes and can be large enough to provide the correct amount of DM annihilation to account for the observed relic abundance.
The freedom to chose one Yukawa to be small to satisfy LFV constraints and the other large enough to give rise to the correct DM relic abundance is a particular feature of Dirac scotogenic models.

\section{Summary and discussion} 
\label{sec:conclusions}

We have examined a scotogenic model with unbroken $U(1)_{B-L}$ gauge symmetry in which neutrino masses are generated at one-loop level, see Fig.~\ref{fig:ScotoLoop}.
Our construction extends the original proposal in Ref. \cite{Leite:2020wjl} by implementing the spontaneous breaking of a global $U(1)_{G}$ symmetry\footnote{It bears common features
  with Ref.~\cite{Calle:2019mxn} but has also important differences, e.g. the Dirac nature of our fermionic dark mediators. }. 
The latter leads to a Goldstone (dubbed Diracon) which can affect the cosmological radiation density $\Delta N_{\text{eff}}$.
The constraint from the Cosmic Microwave Background plus Baryon Acoustic Oscillations is shown in Fig.~\ref{fig:Neff} and implies a multi-TeV  $Z_{\text{BL}}$. 
The interplay of $U(1)_{B-L}$ and $U(1)_{G}$ symmetries ensures cold dark matter (DM) stability and the Dirac nature of neutrinos, forbidding the appearance of Majorana masses.
The diagrams involved in dark matter pair annihilation and direct detection are given in Figs.~\ref{fig:AnnDiag} and \ref{fig:DD}.
Our setup provides a theory framework for $Z_{\text{BL}}$ and $\varphi$-exchange portal dark matter, in which these play a key role in DM annihilation/detection.
Our results on the phenomenology of Dirac fermion singlet scotogenic dark matter are summarized in Fig.~\ref{fig:ZBLNUportalPlots}. 
Dark matter annihilation may proceed via pure $Z_{\text{BL}}$ and $\varphi$-exchange limits, as indicated in Figs.~\ref{fig:ZBLlimit} and \ref{fig:philimit}, respectively.
The magnitude of the Dirac neutrino Yukawa coupling versus DM mass required for the latter is illustrated in Fig.~\ref{fig:ynudm}.
In such $\varphi$-exchange annihilation scenario we also expect a minimum DM mass in the GeV region, as illustrated in Fig.~\ref{fig:FDMlight}.
To sum up, we examined a Dirac scotogenic dark matter framework with gauged $B-L$ and found that consistency between the required neutrino masses and the observed relic dark matter abundance points towards WIMP masses well within reach for upcoming DM experiments.

\black
\begin{acknowledgments}

  Work supported by the Spanish grants FPA2017-85216-P (AEI/FEDER, UE), PROMETEO/2018/165 (Generalitat Valenciana).
  C.A. is supported by NSFC under Grant No. 11975134 and the National Key Research and Development Program of China under Grant No. 2017YFA0402204.
  The work of C.B. has been supported by the FONDECYT grant \textit{Nu Physics} No. 11201240.
  J.L. acknowledges financial support under grants 2017/23027-2 and 2019/04195-7, S\~ao Paulo Research Foundation (FAPESP).
  C.A. thanks the Center of High Energy Physics at Tsinghua University (THU) for its hospitality.
  C.B. and J.L. would like to thank Instituto de F\'isica Corpuscular (CSIC) for the hospitality while part of this work was carried out.

\end{acknowledgments}
\newpage
\bibliographystyle{utphys}
\bibliography{bibliography}
\end{document}